\documentclass[aps,prd,amsmath,superscriptaddress,showpacs,captions=nooneline,twocolumn,preprintnumbers]{revtex4-1} 
\usepackage{graphicx,amsmath,amssymb,bm,color,inputenc,fontenc,mathrsfs}
\usepackage{hyperref} % hypertext links 
\usepackage{epstopdf}
\usepackage{float}
\usepackage[caption = false]{subfig}
\usepackage{comment}
\usepackage{bm} % ----------------------------------------------------------------

\newcommand*{\ket}[1]{\left|{#1}\right\rangle}

\makeatletter
\def\l@subsubsection#1#2{}
\makeatother

\usepackage[dvipsnames,usenames]{xcolor}

\newcommand{\IQUS}{{InQubator for Quantum Simulation (IQuS), Department of Physics, University of Washington, Seattle, WA 98195, USA}}
\newcommand{\UNITN}{{Dipartimento di Fisica, University of Trento, via Sommarive 14, I–38123, Povo, Trento, Italy}}
\newcommand{\TIFPA}{INFN-TIFPA Trento Institute of Fundamental Physics and Applications,  Trento, Italy}

\begin{document}
\title{Entanglement and correlations in fast collective neutrino flavor oscillations}

\author{Alessandro Roggero}
\affiliation{\IQUS}
\affiliation{\UNITN}
\affiliation{\TIFPA}
\email{a.roggero@unitn.it}
\author{Ermal Rrapaj}
\affiliation{Department of Physics, University of California, Berkeley, California 94720, USA}
\email{ermalrrapaj@gmail.com}
\author{Zewei Xiong}
\affiliation{GSI Helmholtzzentrum {f\"ur} Schwerioneneforschung, 64291 Darmstadt, Germany}
\email{z.xiong@gsi.de}

\newcommand{\alert}[1]{\textbf{\color{red}{#1}}}
\renewcommand{\vec}[1]{\boldsymbol{#1}}

\begin{abstract}
  Collective neutrino oscillations play a crucial role in transporting lepton flavor in astrophysical settings like supernovae and neutron star binary merger remnants, which are characterized by large neutrino densities.  In these settings, simulations in the mean-field approximation show that neutrino-neutrino interactions can overtake vacuum oscillations and give rise to fast collective flavor evolution on time-scales $t\propto\mu^{-1}$, with $\mu$ proportional to the local neutrino density. In this work, we study the full out-of-equilibrium flavor dynamics in simple multi-angle geometries displaying fast oscillations in the mean field linear stability analysis. Focusing on simple initial conditions, we analyze the production of pair correlations and entanglement in the complete many-body-dynamics as a function of the number $N$ of neutrinos in the system, for up to thousands of neutrinos.
  Similarly to simpler geometries with only two neutrino beams, we identify three regimes: stable configurations with vanishing flavor oscillations, marginally unstable configurations with evolution occurring on long time scales $\tau\approx\mu^{-1}\sqrt{N}$, and unstable configurations showing flavor evolution on short time scales $\tau\approx\mu^{-1}\log(N)$. We present evidence that these fast collective modes are generated by the same dynamical phase transition which leads to the slow bipolar oscillations, establishing a connection between these two phenomena and explaining the difference in their time scales.
We conclude by discussing a semi-classical approximation which reproduces the entanglement entropy at short to medium time scales and can be potentially useful in situations with more complicated geometries where classical simulation methods starts to become inefficient.
\end{abstract}

\preprint{N3AS-22-006,IQuS@UW-21-023}

\maketitle
%%%%%%%%%%%%%%%%%%%%%%%%%%%%%%%%%
%%%%%%%%%%%%%%%%%%%%%%%%%%%%%%%%%
%%%%%%%%%%%%%%%%%%%%%%%%%%%%%%%%%
%\tableofcontents

%========================================================================================
\section{INTRODUCTION}
\label{sec:introduction}
%========================================================================================

Neutrinos are some of the most abundant particles found in nature, produced during the early universe \cite{Kostelecky:1993, Abazajian:2002, Steigman:2012, Follin:2015}, from stars like the sun during their lifetime~\cite{Davis:1968},
and in copious amounts during core collapse supernovae~\cite{Hirata:1987, Bionta:1987, Alekseev:1987, Qian:1994, Qian:1995, Mirizzi:2015eza}. 
Neutrino flavor conversions, or oscillations, are genuine quantum mechanical phenomena for which a flavor eigenstate is converted to another during
propagation due to it being an admixture of different mass eigenstates.

In core-collapse supernovae (CCSNe) and neutron star merger remnants, neutrinos are responsible for both reinvigorating a stalled shock-wave and controlling the conditions for nucleosynthesis in the ejected material~\cite{Hoffman:1997,Janka:2012,Winteler:2012,Wanajo:2014}. In these environments neutrino flavor evolution is substantially modified by the presence of neutrino-neutrino scattering processes which can lead to self-sustained collective flavor oscillations~\cite{Savage:1991,Pantaleone:1992,Pantaleone:1992v2,Pastor:2002,Balantekin:2005,Fuller:2006,Duan:2006,Friedland:2010,Wu:2017}. Since neutrinos in supernovae are emitted with fluxes and spectra that are strongly flavor dependent~\cite{Janka:2012}, the presence of collective flavor oscillations could then lead to important effects~\cite{Fogli:2007,Qian:1993,Qian:1995v2,Duan:2007,Dasgupta:2008cd,Gava:2009,Duan:2009,Balantekin:2009dy,
Dasgupta:2010cd,Duan:2010,Raffelt:2010,Raffelt:2011,Duan:2011,Cherry:2012zw}. Neutrino-neutrino scattering, being between particles of the same type, is of a different nature than neutrino-matter scattering, and gives rise to forward scattering terms in the many-body Hamiltonian which 
contribute to oscillations \cite{Pantaleone:1992, Pantaleone:1992-2}. These terms are dependent only on the angle between neutrinos and couple neutrinos of different energies making flavor
evolution a rather intricate many-body problem. Thanks to the adoption of a mean-field approximation, a rich phenomenology of collective neutrino modes have been identified (see~\cite{Duan2010,Chakraborty2016b} for reviews). In particular two main classes of collective modes have been categorized as the slow and fast modes of flavor instability based on the triggering mechanism and the typical length scale of the flavor transition.
Slow modes are due to the interference of the vacuum flavor mixing and neutrino-neutrino self-induced forward scattering. The respective conversion rate is $\sim \sqrt{\omega\mu}$, where $\omega = \Delta m^2/2E_\nu$ is the vacuum oscillation frequency for neutrinos of energy $E_\nu$ with mass square difference $\Delta m^2$, and $\mu= \sqrt{2} G_F \rho_\nu$ indicates the magnitude of self-induced effective potential with Fermi constant $G_F$ and neutrino number density $\rho_\nu$.
Slow flavor evolution typically shows a bipolar behavior in terms of the flavor survival probability and usually results in drastic splitting of neutrino spectra \cite{Duan:2007, Raffelt:2007, Duan2010, Chakraborty2016b}.
Fast flavor conversions can occur even in the absence of vacuum mixing since they are triggered by non-trivial angular distributions and the consequent flavor evolution has a strong angular dependence.
The associated flavor conversion rate is $\sim \mu$, much faster than the slow mode when the neutrino number density $\rho_\nu$ is high and $\mu\gg \omega$ as, for example, near the proto-neutron star of CCSNe or the hyper-massive star of merger remnants  \cite{Sawyer:2005, Sawyer:2015, Capozzi:2017, Izaguirre:2017, Yi:2019, Johns:2020, Tamborra:2020, Xiong:2021s}.

In this work we study collective oscillations of two active neutrino flavors, under only the influence of the Hamiltonian induced by neutrino-neutrino interactions.
We assume a simplified scenario of electron neutrinos $\nu_e$, and an additional flavor which can be considered as a superposition of tau and muon 
neutrinos denoted by $\nu_x$, with no vacuum mixing (or high neutrino density) and only focus on the effects of neutrino forward scattering. To simplify the treatment, we consider a three beam setup as explained in section~\ref{sec:three_beam}.
With $N_f$ flavors and neglecting momentum-changing interactions, the many-body Hamiltonian can be formulated in terms of $SU(N_f)$ operators acting on the flavor state of neutrinos. This approach is particularly useful for studying many-body effects. In section~\ref{sec:lin_stab} we perform a linear stability analysis in the mean field approximation to determine which configurations are unstable under perturbations, and proceed to explain the many-body methods used in the work in section~\ref{sec:methods}. The results for the flavor evolution for the various setup and increasing particle number are summarized in section~\ref{sec:flavor_evolution_results}. In section~\ref{sec:entanglement_correlations} we focus on the dynamical creation of entanglement entropy and correlations from the initial mean field wavefunction.
The findings are summarized and conclusions are drawn in section~\ref{sec:summary}.  

\section{Three beam geometry and Hamiltonian}
\label{sec:three_beam}
As our focus here is on the many-body effects, we consider only the flavor evolution of neutrinos under $\nu-\nu$ forward scattering and ignore the vacuum term or scattering with matter. In studies of collective and fast neutrino flavor oscillations, this is a common choice as the flavor instability is assumed to originate from this part of the total Hamiltonian, with an initial ``seed'' from the other terms~\cite{Sawyer:2015,Izaguirre:2017,Capozzi:2017,Yi:2019,Bhattacharyya:2020,Padilla_Gay:2021,Bhattacharyya:2021,Wu:2021}. This work is the first attempt at uncovering the neutrino-neutrino correlations and quantum entanglement using the complete many-body treatment of this dynamics under the influence of multi-angle effects. To study the large particle number limit, we assume a constant neutrino density $\rho_{\nu}$, and the system to be comprised of several neutrino beams (directions). Each beam contains many neutrinos with momenta aligned to each other~\footnote{Due to the fermionic anti-symmetry, we assume neutrinos propagate in almost parallel direction but neglect the effects of misalignment.}.  
Accounting only for forward scattering, the Hamiltonian governing flavor evolution can thus by expressed in the following from~\cite{Pehlivan:2011}
\begin{equation}
 H = \frac{\mu}{ N} \sum_{i \neq j}^{N}(1-c_{ij})\vec{J}_i\cdot\vec{J}_j
 \label{eq:Hamiltonian}
\end{equation}
with $N$ the total particle number and $c_{ij}=\cos(\theta_{ij})$ the cosine of the angle between the momenta of neutrinos $i$ and $j$. The interactions between neutrinos propagating in parallel directions therefore vanishes. The coupling constant $\mu=\sqrt{2}G_F\rho_\nu$ depends on both Fermi's constant $G_F$ and the local neutrino density $\rho_\nu$. Here we work in the approximation where neutrinos have only two possible flavors and their state can be specified using a two component isospin degree of freedom. The single particle operators acting on these flavor states form an $SU(2)$ algebra and can be expressed as
\begin{equation}
\vec{J}_i = \frac{1}{2}\left(\sigma^x_i,\sigma^y_i,\sigma^z_i\right)\;,
\end{equation}
with $\sigma^k_i$ the $k$-th Pauli matrix acting on the $i$-th particle. We can also define beam operators as
\begin{equation}
\vec{J}_{A_i} = \sum_{k\in A_i} \vec{J}_i\;,
\end{equation}
where the sum runs over the $N_{A_i}$ particles belonging to the $i$-th beam. Since $[H,\vec{J}^2_{A_i}]=0$, the total flavor isospin of each beam is conserved and, for initial states that are eigenstates of $\vec{J}^2_{A_i}$, we can express the Hamiltonian in terms of beam operators as follows
\begin{equation}
 H =  \frac{2\mu}{ N} \sum_{i<j}^{n}(1-c_{A_i A_j})\vec{J}_{A_i}\cdot\vec{J}_{A_j}\;,
 \label{eq:beam_Hamiltonian}
\end{equation}
where $n$ is the number of beams  and have neglected irrelevant additive constant terms. This system has many symmetries worth pointing out.
In addition to the individual $\vec{J}^2_{A_i}$ being conserved, the total angular momentum commutes with the Hamiltonian ($[\vec{J}^2,H_{ABC}]=0$), and $\langle \vec{J}^2 \rangle_{\Psi_{1,2}} = N^2/36  + N/2$ is a constant of the motion. In addition, $[\vec{J},H]=0$ and $\langle \vec{J} \rangle_{\Psi_{1,2}} = (0,0,N/6)$ is a conserved quantity, as well.

\begin{figure}[t]
\includegraphics[scale=0.45]{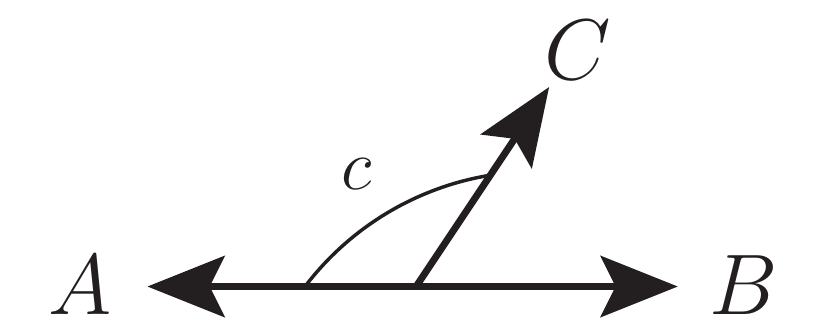}
\caption{Beams $A$ and $B$ are antiparallel, and beam $C$ forms an angle $\theta_{AC}$ with beam $A$.}
\label{fig:threebeams} 
\end{figure}

For simplicity, we will take $n=3$ beams and assume an equal number of neutrinos in each beam with $N_{A_i}=N/3$. We further consider the direction of propagation in these three beams to lie on a plane and that two of them are antiparallel. This simple angular configuration is shown in Fig.~\ref{fig:threebeams} and is parametrized by a single angle $\theta_{AC}$.

The Hamiltonian then becomes (see. Appendix~\ref{app:3beams})
\begin{equation}
\begin{split}
H_{ABC} &= \frac{4\mu}{N}\vec{J}_A\cdot\vec{J}_B+\frac{2\mu}{N}(1-c)\vec{J}_A\cdot\vec{J}_C\\
&+\frac{2\mu}{N}(1+c)\vec{J}_B\cdot\vec{J}_C\;,
\end{split}
\label{eq:Habc}
\end{equation}
where we introduced $c=\cos(\theta_{AC})$. This Hamiltonian is invariant for a global $SU(2)$ rotation and take as convention the $z$ axis to be flavor axis. We consider two distinct initial configurations diagonal in flavor
\begin{enumerate}
  \item $\ket{\Psi_{1}(0)} = \ket{\uparrow}^{\otimes N_A}\otimes\ket{\uparrow}^{\otimes N_B}\otimes\ket{\downarrow}^{\otimes N_C}$,
  \item $\ket{\Psi_{2}(0)} = \ket{\uparrow}^{\otimes N_A}\otimes\ket{\downarrow}^{\otimes N_B}\otimes\ket{\uparrow}^{\otimes N_C}$,
 \end{enumerate}
 and $N_A=N_B=N_C=N/3$. In the text we will refer to these initial conditions as setup I and setup II. The convention we use throughout is that electron flavor is associated with an up-spin and the heavy lepton flavor $\nu_x$ with a down-spin. This rather special choice of initial states was motivated by the fact that a mean-field treatment of their time propagation will result in no flavor evolution and therefore any flavor dynamics is inherently a many-body effect. Adding a small off-diagonal component leads to an evolving mean-field solution which we will use to characterize the stability of the resulting equation of motion.

\section{Mean-field linear stability analysis}
\label{sec:lin_stab}
In the mean-field approximation it is commonly assumed that the correlation between any two neutrinos are negligible: $\langle \mathcal O_i \mathcal O_j \rangle = \langle \mathcal O_i \rangle \langle \mathcal O_j \rangle$ where $i$ and $j$ are indices for different neutrinos. 
Therefore, the time evolution for each neutrino is rewritten as
\begin{equation}
\begin{split}
    \partial_t \langle \vec{J}_i \rangle =
    \vec{H}_\mathrm{MF} \times \langle \vec{J}_i \rangle
    = \frac{2\mu}{N} \sum_{j \neq i}^{N}(1-c_{ij}) \langle \vec{J}_j \rangle \times \langle \vec{J}_i \rangle.
\end{split}
\end{equation}
This mean field approximation can be expected to hold in the limit of large quantum numbers and therefore it is convenient to formulate the evolution equations treating all the neutrinos in a beam at the same time. One can then define a normalized polarization vector for each beam as $\vec{\mathcal P}_{A_i} = \sum_{i\in A_i} \langle \vec{J}_i \rangle / (N_{A_i}/2)$ and rewrite the equation of motions in terms of $\vec{\mathcal P}_{A_i}$. For our three beam setup we have then
\begin{align}
    \partial_t \vec{\mathcal P}_A & = \frac{\mu}{N} \left[ 2 N_B \vec{\mathcal P}_B + (1-c) N_C \vec{\mathcal P}_C \right] \times \vec{\mathcal P}_A \nonumber\\
    \partial_t \vec{\mathcal P}_B & = \frac{\mu}{N} \left[ 2 N_A \vec{\mathcal P}_A + (1+c) N_C \vec{\mathcal P}_C \right] \times \vec{\mathcal P}_B \nonumber\\
    \partial_t \vec{\mathcal P}_C & = \frac{\mu}{N} \left[ (1-c) N_A \vec{\mathcal P}_A + (1+c) N_B \vec{\mathcal P}_B \right] \times \vec{\mathcal P}_C.
\end{align}

The instability of the neutrino gas can be diagnosed by analyzing the stability of these differential equations to small perturbations. In the neutrino case, we will then assume that the third component $\mathcal P^z_{A_i}$ is dominant and linearize the mean-field equations of motion (EOM) in terms of the perturbation away from the flavor axis.
Given $N_A=N_B=N_C=N/3$, the linearized EOM for the off-diagonal component $\mathcal{S}_{A_i}\equiv \mathcal{P}^x_{A_i} -i \mathcal{P}^y_{A_i}$ reads
\begin{widetext}
\begin{equation}
    \partial_t \begin{pmatrix} \mathcal{S}_A \\ \mathcal{S}_B \\ \mathcal{S}_C \end{pmatrix}
    = M_\mathrm{LMF}
    \begin{pmatrix} \mathcal{S}_A \\ \mathcal{S}_B \\ \mathcal{S}_C \end{pmatrix}
    = \frac{\mu}{3} \begin{pmatrix}
        2 {\mathcal P}^z_B + (1-c) {\mathcal P}^z_C & -2 \mathcal{P}^z_A & -(1-c) \mathcal{P}^z_A \\
        -2 {\mathcal P}^z_B & 2 {\mathcal P}^z_A + (1+c) {\mathcal P}^z_C & -(1+c) {\mathcal P}^z_B \\
        -(1-c) {\mathcal P}^z_C & -(1+c) {\mathcal P}^z_C & (1-c) {\mathcal P}^z_A + (1+c) {\mathcal P}^z_B
    \end{pmatrix}
    \begin{pmatrix} \mathcal{S}_A \\ \mathcal{S}_B \\ \mathcal{S}_C \end{pmatrix}.
\end{equation}
\end{widetext}
The unstable mode of the neutrino gas can be found by parametrizing the time-dependence of the off-diagonal component as $\mathcal{S}_{A_i} = Q_{A_i} e^{-i\Omega t}$ and solving the collective oscillation frequency $\Omega$ as the eigenvalues of the matrix $M_\mathrm{LMF}$.
Any eigenvalues with positive imaginary components imply the existence of modes with exponentially growing amplitudes, which have been associated with the appearance of fast flavor conversion~\cite{Banerjee:2011,Izaguirre:2017}.

The value of $\mathcal P^z_{A_i}$ can be either $+1$ or $-1$ and depends on the choice of initial conditions.
For the state $\ket{\Psi_1}$ from setup I, the eigenvalue equation gives
\begin{equation}
    \Omega \left[ 9 \left(\frac{\Omega}{\mu}\right)^2 - 12 \frac{\Omega}{\mu} + 3 + c^2  \right]  = 0.
\end{equation}
Since the quadratic discriminant $\Delta = 36( 1 - c^2) $ is non-negative, there is no flavor instability at the mean-field level. For setup II, we have
\begin{equation}
    \Omega \left[ 9 \left(\frac{\Omega}{\mu}\right)^2 - 6 (1-c) \frac{\Omega}{\mu} + 1 - 4c - c^2 \right] = 0.
\end{equation}
Since $\Delta = 72 c(c+1) $, when $-1<c<0$, there is flavor instability.
The unstable solution is
\begin{equation}
    \Omega = \frac{ (1-c) \pm \sqrt{2c(c+1)}}{3} \mu.
\end{equation}
When $c=-1/2$, the growth rate reaches the maximum value $\sqrt{2} \mu/6$.
The value of $\Omega$ for the unstable mode in setup II can be plugged back into the linearized EOM to obtain the following relations of the corresponding eigenvector compared to that of the $\nu_x$ beam: 
\begin{equation}
\begin{split}
    \frac{|Q_A|^2}{|Q_B|^2} &= \frac{1+c}{1-c}, \\
    \frac{|Q_C|^2}{|Q_B|^2} = 1-&\frac{|Q_A|^2}{|Q_B|^2}
    = \frac{2 c}{c-1}.
\end{split}
    \label{eq:linear_analysis_eigenvector}
\end{equation}
The transverse components are associated with flavor transitions in each beam in the linear regime.
A larger value for the amplitude $|Q_{A_i}|^2$ leads to a higher change of flavor content in the corresponding beam
\begin{equation}
\begin{split}
\left|\mathcal P^z_{A_i}(t)-\mathcal P^z_{A_i}(0)\right| &\approx \frac{|S_{A_i}(t)|^2}{2} \\
&\approx \frac{|Q_{A_i}|^2}{2} \cdot e^{2 \text{Im}(\Omega) t}\;.   
\end{split}
\end{equation}

When the angular parameter $c$ approaches 0, $|Q_C|^2$ is smaller than $|Q_A|^2$ and the flavor conversion is primarily associated with beam $A$ rather than $C$.
On the other hand, beam $C$ has more flavor transition when $c$ approaches -1.
While the above relations may not be valid for long time scales, they can describe which $\nu_e$ beam is mostly associated with the flavor conversion when the system transits from the linear to the non-linear regime.

 \section{Methods}
 \label{sec:methods}
 In this section we briefly describe the strategy we employ to perform simulations of the three-beam model from Eq.~\eqref{eq:Habc} with systems up to $N=2700$. These system sizes are much larger than what would be possible using the tensor network methods employed in previous works~\cite{Roggero:2021,PhysRevD.104.123023,Cervia:2022}. Efficient simulations are made possible through an effective use of the angular momentum representation (see~\cite{Xiong:2021} for the general method and~\cite{martin2021classical} for an application to a two-beam model).

To give a concrete example, the initial wave function for setup I in this basis is written as
\begin{equation}
    \ket{\Psi(0)} 
    = \ket{j_A, m_A} \otimes \ket{j_B, m_B} \otimes \ket{j_C, m_C},
    \label{eq:initial_wavefunction_angular_momentum}
\end{equation}
where $j_A=m_A=N_A/2$, $j_B=m_B=N_B/2$, and $j_C=-m_C=N_C/2$.
The total flavor isospin of each beam, $j_A$, $j_B$, or $j_C$, is conserved and can be determined from the initial condition.
A simplified many-body notation can be introduced as $\ket{\Psi} = \ket{m_A, m_B}$ with only two degrees of freedom, where $m_C$ is determined by $m_A$ and $m_B$ given that the total projection of flavor isospin, $m_A+m_B+m_C$, is conserved.
The evolving state is then a linear combination of states with all possible $m_A$ and $m_B$,
\begin{equation}
    \ket{\Psi(t)} = \sum_{m_A, m_B} a_{m_A, m_B}(t) \ket{m_A, m_B}.
    \label{eq:total_wavefunction_2beam}
\end{equation}

We solve the time evolution for the amplitudes of many-body states described above (for more details see appendix~\ref{app:angular_momentum_method}).
Once the amplitudes are known, the observables such as polarization and entanglement entropy can be calculated.
The projection of flavor isospin for each beam is
$
    \langle J^z_{A_i} \rangle
    = \sum_{m_A, m_B} m_{A_i} |a_{m_A, m_B}|^2
$.
Pair correlations are
$
    \langle J^z_{A_i} J^z_{A_j} \rangle = \sum_{m_A, m_B} m_{A_i} m_{A_j} |a_{m_A, m_B}|^2
$.
The correlations along the other two directions in flavor space are
$
    \langle J^x_{A_i} J^x_{A_j} \rangle = \langle J^y_{A_i} J^y_{A_j} \rangle = \langle J^+_{A_i} J^-_{A_j} + J^-_{A_i} J^+_{A_j} \rangle /4 ,
$
where $J^\pm_{A_i}=J^x_{A_i} \pm i J^y_{A_i}$.
Note that the terms $\langle J_{A_i}^+J_{A_j}^+ \rangle$ and $\langle J_{A_i}^-J_{A_j}^- \rangle$ are both zero because the net flavor isospin, $m_A+m_B+m_C$, is a conserved quantity for the system, and $J_{A_i}^\pm J_{A_j}^\pm |m_A+m_B+m_C\rangle \propto |m_A+m_B+m_C\pm 2\rangle$, leads to violations of this quantity.
Detailed expressions in terms of amplitudes can be found in appendix~\ref{app:angular_momentum_method}.

R\'{e}nyi entropy is an important measure for the entanglement in a subsystem.
For a general multi-qubit system that is divided into two subsystems, $\mathrm{I}$ and $\mathrm{II}$, the R\'{e}nyi entropy of subsystem $\mathrm{I}$ is defined as
\begin{equation}
    \mathcal R_{\alpha, \mathrm{I}} = \frac{1}{1-\alpha} \log_2 [\mathrm{Tr}(\rho_\mathrm{I}^\alpha)],
    \label{eq:renyi_entropy_gen}
\end{equation}
where $\rho_I=\mathrm{Tr}_\mathrm{II}(\rho)$ is the reduced density matrix of subsystem $\mathrm{I}$.
As an example, the R\'{e}nyi entropy of beam $A$ in setup I is given as, 
\begin{equation}
    \mathcal R_{\alpha, A}
    = \frac{1}{1-\alpha} \log_2 \left[ \sum_{m_A=-N_A/2}^{N_A/2} \left( \sum_{m_B} |a_{m_A, m_B}|^2 \right)^\alpha \right].
    \label{eq:R_3beam}
\end{equation}
The Von Neumann entropy can be expressed as R\'{e}nyi entropy in the limit of $\alpha\to 1$
\begin{equation}
     S_\mathrm{I} = \lim_{\alpha\to 1} \mathcal R_{\alpha, \mathrm{I}} = -\mathrm{Tr}[\rho_{\mathrm{I}}\log_2(\rho_{\mathrm{I}})],
    \label{eq:von_neumann_entropy_gen}
\end{equation}
or more explicitly in terms of amplitudes
\begin{equation}
     S_{A} =\!\!\!\!
    \sum_{m_A=-N_A/2}^{N_A/2} \left[ \left( \sum_{m_B} |a_{m_A, m_B}|^2 \right) \log_2 \left( \sum_{m_B} |a_{m_A, m_B}|^2 \right) \right].
\end{equation}

Because setup II can be obtained from setup I by exchanging configurations between beam $B$ and $C$,  all quantities defined can be modified accordingly and not explicitly listed here.
 
 \section{Results for flavor evolution}
 \label{sec:flavor_evolution_results}

 In the following we will first focus on studying the flavor evolution for three beam models in the two setups and show their qualitative differences.
In particular, we will compute the survival probability, or persistence, $P_i(t)$ of a representative neutrino in each beam. This can be defined explicitly in terms of the $\vec{J}_i$ operators as
 \begin{equation}
 \label{eq:surv_prob}
 P_{ik}(t) = \frac{1}{2} + \frac{s_{ik}}{N} \langle \Psi_k(t)\lvert J^z_i \rvert\Psi_k(t)\rangle\;,
 \end{equation}
 with $k=1,2$ denoting to employed initial state and the constant $s_{ik}$ defined as
 \begin{equation}
 s_{ik} = \text{sign}\left[\langle \Psi_k(0)\lvert J^z_i \rvert\Psi_k(0)\rangle\right]\;,
 \end{equation}
 to ensure $P_i(t=0)=1$ for all neutrinos. For ease of notation, in the following we will denote expectation values at time $t$ as $\langle\cdot\rangle(t)$ dropping the index $k$ indicating the initial condition when no risk of confusion arises.
 
 \subsection{Setup I}
 \label{sec:psi1}
 
 The initial wavefunction for setup I is the product state 
 \begin{equation}
 \ket{\Psi_{1}(0)} = \ket{\uparrow}^{\otimes N/3}\otimes\ket{\uparrow}^{\otimes N/3}\otimes\ket{\downarrow}^{\otimes N/3}\;,    
 \end{equation}
 with equal populations in the three beams. This initial state is symmetric under the exchange $A\Leftrightarrow B$ and the Hamiltonian in Eq.~\eqref{eq:Habc} remains invariant under this permutation if we also exchange $c\Leftrightarrow-c$. In our study of this system we will therefore limit the discussion to positive values of the angular parameter $c$.

 The case with $c=0$ is special as for this geometry the total spin $\vec{J}^2_{AB}=(\vec{J}_A+\vec{J}_B)^2$ is also conserved and the Hamiltonian takes the simpler form
 \begin{equation}
  \begin{split}
  \label{eq:czero_setup1}
   H_{ABC} (c=0) =& \frac{4\mu}{N}\vec{J}_A\cdot\vec{J}_B+\frac{2\mu}{N}(\vec{J}_A+\vec{J}_B)\cdot\vec{J}_C \\
   =& N \frac{\mu}{9} + \frac{2 \mu}{N} \left(\vec{J}_{AB}\cdot\vec{J}_C \right)\;.
  \end{split}
 \end{equation}
 We see then, that up to an overall constant, this case reduces to a two-beam model with unequal population numbers. An exact analytical solution for this scenario was already discussed in Ref.~\cite{Friedland:2006} where it was shown that flavor oscillations are present with an amplitude decaying as a polynomial in the population difference $|N_{AB}-N_C|=N/3$. This case recovers the mean field solution qualitatively, which does not show flavor oscillations, in the large system size limit. Our many-body simulations show that this behavior is actually generic for any value of the angular factor $c\neq1$.
 The case $c=1$ is in fact also special as the total spin $\vec{J}^2_{AC}=(\vec{J}_A+\vec{J}_C)^2$ remains conserved and the Hamiltonian becomes
 \begin{equation}
  \begin{split}
  \label{eq:c1_s1}
   H_{ABC} (c=1) =& \frac{4\mu}{N}\vec{J}_{AC}\cdot\vec{J}_B.
  \end{split}
 \end{equation}
 The crucial difference is however that now the two beams $A$ and $C$ have opposite flavor polarization and their total spin is instead $\langle\vec{J}^2_{AC}\rangle=N/3$. A similar situation was also considered in Ref.~\cite{Friedland:2006} but the beam had maximal $\langle\vec{J}^2_{AC}\rangle$ and $\langle J^z_{AC}\rangle=0$ initially (ie. fully polarized in the $xy$-plane). As the behavior in our case for $c=1$ is markedly different from the other ones, we first discuss the case $c\neq1$ and move to $c=1$ near the end of this section.
 
\begin{figure}[t]
  \centering
\includegraphics[width=0.48\textwidth]{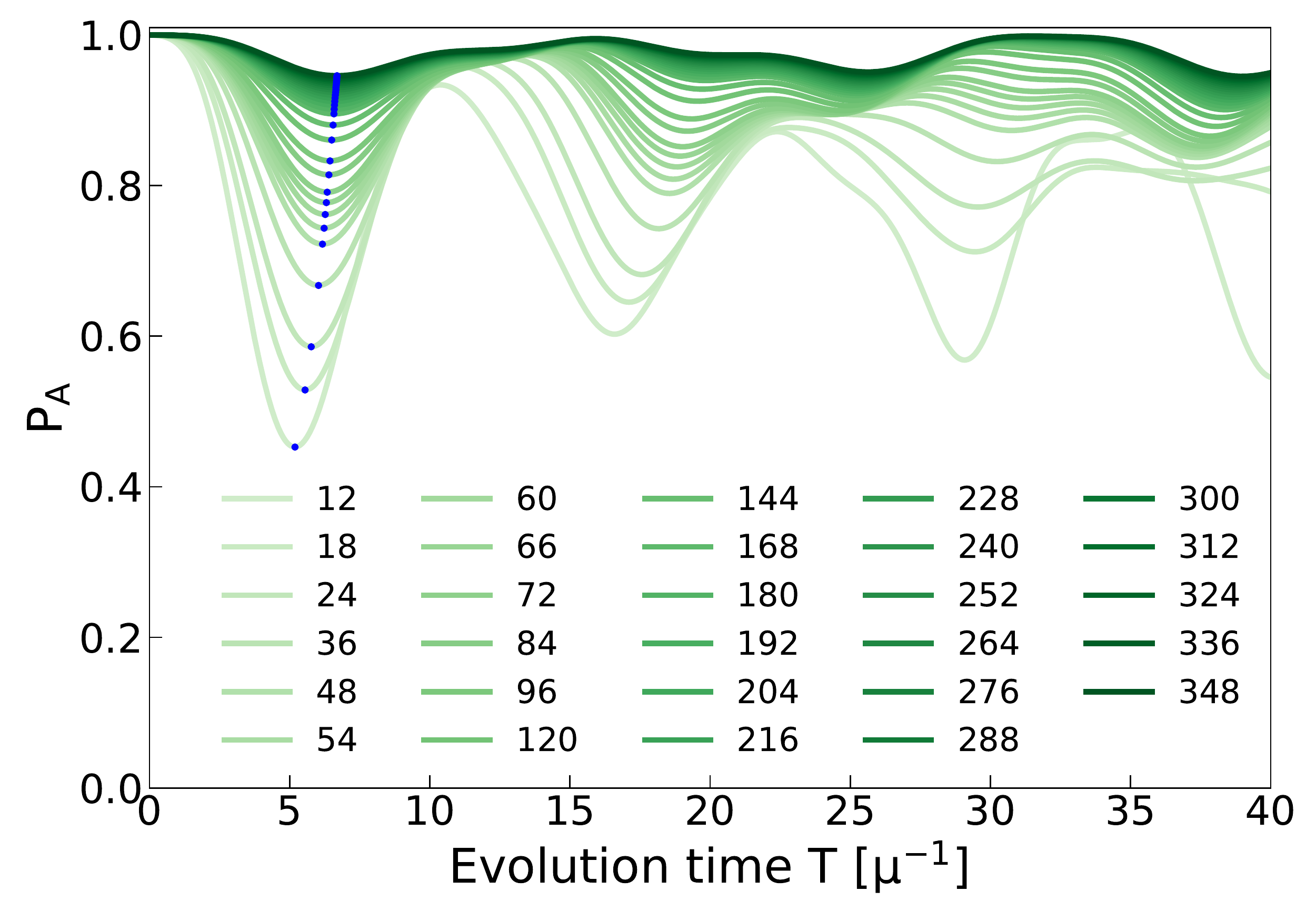}
  \caption{Time evolution of the survival probability starting in the first beam from the initial state of Setup I and taking $c=0.5$ for a large selection of system sizes (green solid curves, darker colors indicate progressively larges values of $N$). With blue dots we also show the location of the first minimum
 }
 \label{fig:3b112_beam0_c0p5_P}
 \end{figure}

  We start by looking at the qualitative behavior of the flavor survival probability for $c=0.5$. In Fig.~\ref{fig:3b112_beam0_c0p5_P} we show results for the evolution of the survival probability $P_A(t)$ in the first beam as a function of the evolution time and for a variety of system sizes ranging from $N=12$ to $N=348$ (indicated with increasingly darker colors for larger systems). The qualitative evolution remains the same for other values of $c\neq1$ and for different beams.
 
\begin{figure}[b]
 \includegraphics[width=0.48\textwidth]{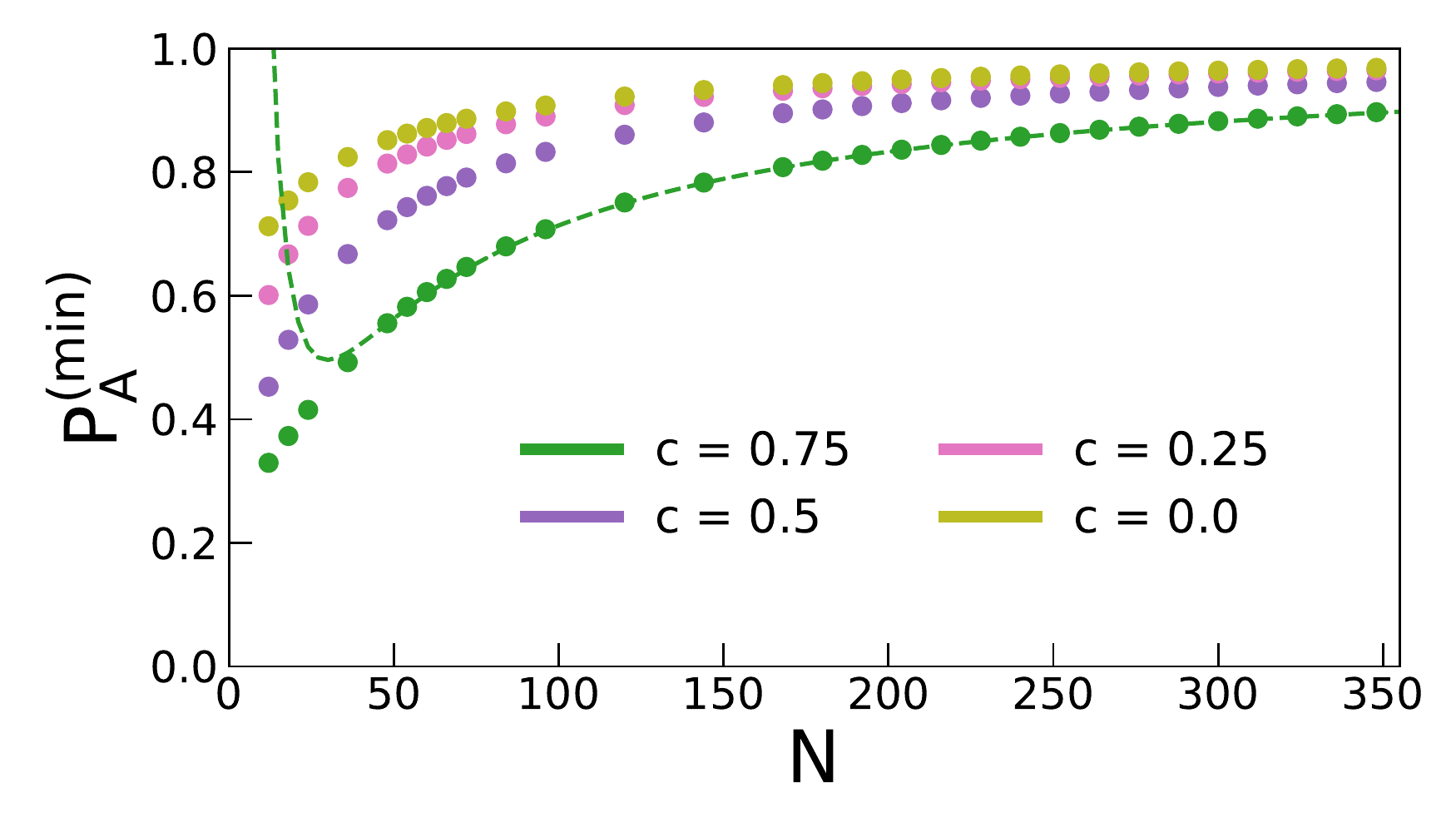}
\caption{Evolution with system size $N$ of the value at the first minimum of the survival probability for beam A using the initial wavefunction $\ket{\Psi_1}$ and various cosine values. The dashed green line corresponds to the best fit for $c=0.75$ using the parameterization from Eq.~\eqref{eq:pmin_ansatz}. }
 \label{fig:3b112_polarization_min}
 \end{figure}

 In order to more easily track the evolution of the amplitude of flavor oscillations in the large $N$ limit, we also show in Fig.~\ref{fig:3b112_beam0_c0p5_P} the location of the first minimum of the survival probability using blue dots. 
  In the following we will indicate the value reached at the first minimum of the survival probability in beam $A_i$ as $\rm P_{A_i}^{(min)}$. The results for beam $A$ and different values of the angular distribution parameterized by $c$ are shown in Fig.~\ref{fig:3b112_polarization_min}. We find that in all cases the survival probability converges to $1$ in the large system size limit. For large but finite $N\gtrapprox50$ the scaling with $N$ is well reproduced by the simple ansatz
\begin{equation}
\label{eq:pmin_ansatz}
\rm P^{(min)}_{A_i}(N) \approx 1 - \frac{a}{N}\left(1-\frac{b}{\sqrt{N}}\right)\;,
\end{equation}
with $b=\mathcal{O}(1)$ and $a$ increasing with the angular parameter $c$ from $a\approx13$ at $c=0$ to $a\approx45$ at $c=0.75$. Due to the relatively limited maximum system size considered here, we found the correction term parameterized by $b$ to be important for all angular distributions even though its contribution will vanish in the thermodynamic limit. In Fig.~\ref{fig:3b112_polarization_min} we show the fit performed using Eq.~\eqref{eq:pmin_ansatz} for the case $c=0.75$ as a green dashed line.
 
 In terms of the expectation value of the spin operators, the scaling from Eq.~\eqref{eq:pmin_ansatz} indicates that in the many-body evolution the expectation value of $\langle J_i^z\rangle$ deviates from its initial value $\pm N_i/2$ only by a constant factor
 \begin{equation}
\left|\langle J_i^z\rangle \right|\gtrapprox \frac{N_i}{2}-a\frac{N_i}{2N}\left(1-\frac{b}{\sqrt{N}}\right)\;,
 \end{equation}
 and the fractional change measured by the z component of the polarization vectors $\vec{\mathcal{P}}_{A_i}$, defined in Sec.~\ref{sec:lin_stab} and used in the mean-field approximation, vanishes for large systems. A similar pattern can also be observed in the other two beams. However, in the second beam we noticed a transient behavior where the first minimum transitions to a stationary point as the system size increases, and the initial second minimum becomes the first one after $N\approx 100$. Regardless, Eq.~\eqref{eq:pmin_ansatz} remains valid also for this beam for large enough system values (after the transition from first minimum to stationary point).  
 
 The time scale to reach the first minimum of the survival probability seems to converge to a constant in the large system size limit agreeing with the expectations from the study in Ref.~\cite{Friedland:2006} which were obtained for $c=0$. 
 
 As mentioned above, the case $c=1$ is peculiar in that the total spin of beams $A$ and $C$ is conserved and kept for all times at a small value $\langle J^2_{AC}\rangle=N/3$ comparable with the size of quantum fluctuations in the total spin $\langle (J^x)^2\rangle=\langle (J^y)^2\rangle=N/4$. Contrary to the previous cases, this allows for quantum fluctuations to drive flavor evolution in a similar way as in the simpler two beam model studied in a previous works (see~\cite{Roggero:2021,PhysRevD.104.123023,Friedland:2003eh,Friedland:2006}). Interestingly however, in this case beams $A$ and $C$ are only coupled trough their interaction with beam $B$.
 
\begin{figure}[t!]
\subfloat[]{\includegraphics[width=0.48\textwidth]{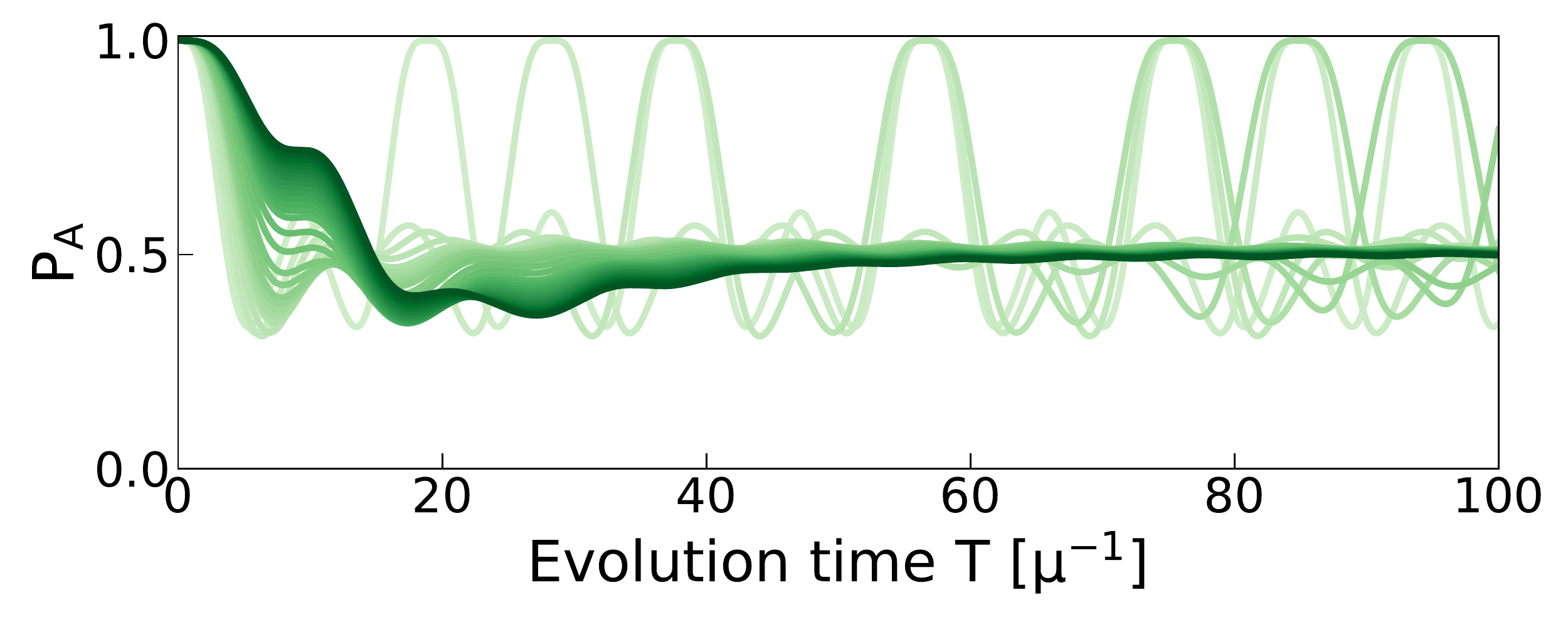}}\\
\subfloat[]{\includegraphics[width=0.48\textwidth]{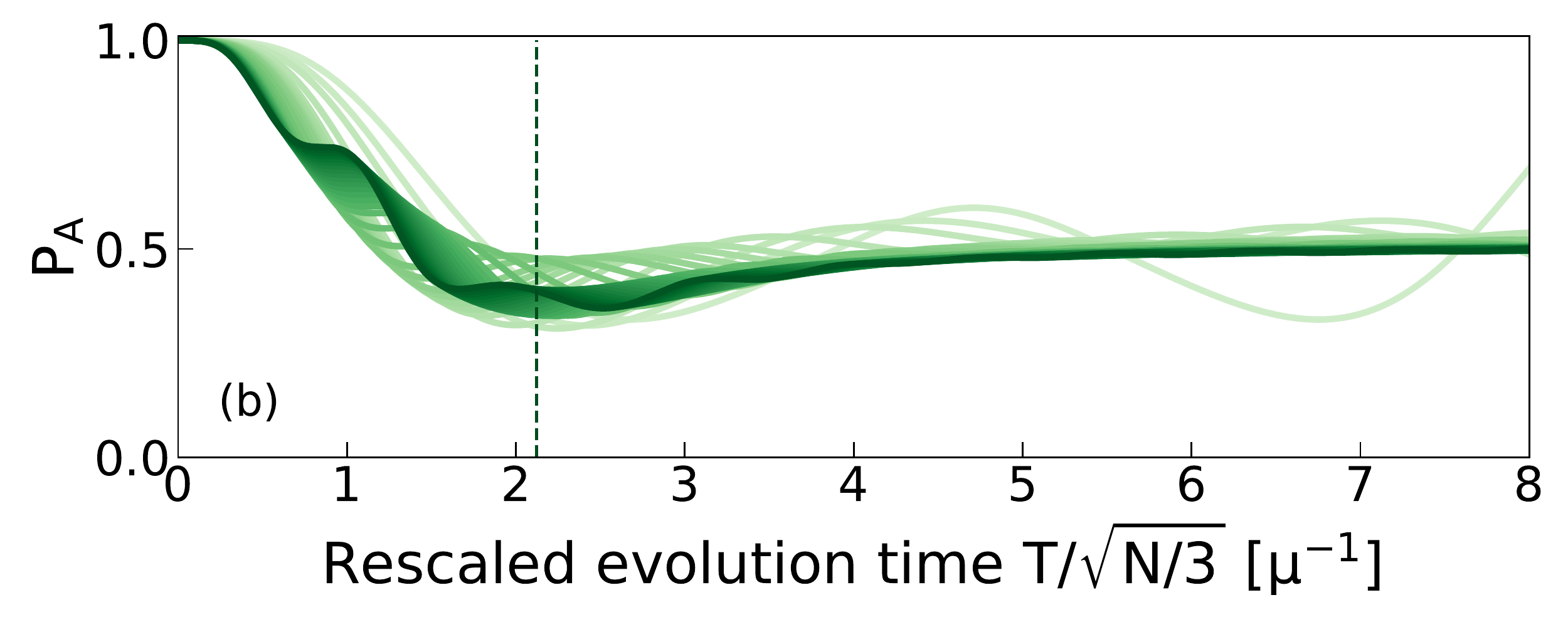}}
\caption{Time evolution of the survival probability in beam $A$ for the initial state in Setup I, $c=1$ and the same set of system sizes used for Fig.~\ref{fig:3b112_beam0_c0p5_P} (darker colors indicate larger values of $N$). Panel $(a)$ shows directly the time evolution while panel $(b)$ uses a rescaled time variable $T/\sqrt{N/3}$ to highlight the system size dependence. The vertical dashed line in panel (b) indicates the time scale $\tau_{AC}$ expected for a two-beam model.}
\label{fig:3b112_beam_0_c1_P}
\end{figure}
 
 We show the result of our simulation for the survival probability in beam $A$ for this case in Fig.~\ref{fig:3b112_beam_0_c1_P}. The behavior of beam $C$ is the same while beam $B$ shows little flavor conversion similarly to the results shown in Fig.~\ref{fig:3b112_beam0_c0p5_P}. The top panel shows the evolution of $P_A$ as a function of total time $T$. This is in marked contrast with the results seen above for $c\neq1$: the survival probability converges to $0.5$ (full mixing) for large times displaying oscillations whose amplitude decays away in the limit of large systems (darker curves in the plot). In order to display more clearly the system size dependence of the time scale to reach the plateau, we also show in the bottom panel of Fig.~\ref{fig:3b112_beam_0_c1_P} the same data but as a function of the rescaled time variable $T/\sqrt{N/3}$. These results clearly indicate a decaying time scale $\tau\approx2\mu^{-1}\sqrt{N/3}$ to reach equilibration at $P_A=0.5$. Apart from the superimposed oscillations, this is remarkably similar to the behavior shown by a two-beam system, initialized in opposite flavor states. In order to isolate the effective Hamiltonian for the two beams we rewrite
 \begin{equation}
  H_{ABC} (c=1) = \frac{2\mu}{N}\left(\vec{J}^2-\vec{J}^2_{AC}\right)-\mu\frac{N+6}{18}\;,
 \end{equation}
 with $\vec{J}^2=\vec{J}^2_{ABC}$ the total angular momentum.
The two angular momentum operators commute and can then be applied sequentially. The contribution proportional to the total angular momentum is proportional to $H_{ABC}(c=0)$ in Eq.~\eqref{eq:czero_setup1} (apart from a constant factor) and, as shown above, does not lead to stable oscillations in the $N\to\infty$ limit. The large $N$ evolution of the configuration with $c=1$ is then captured by the effective two-beam Hamiltonian
 \begin{equation}
 \label{eq:twob_ac}
 H^{\rm 2 Beams}_{AC} = -\frac{2\mu}{N} J^2_{AC} = -\frac{4}{3} \frac{\mu}{2N/3} J^2_{AC}\;.
 \end{equation}
 Using the results from Ref.~\cite{Roggero:2021}, the time scale obtained to reach the minimum of the survival probability would be $\tau_{AC}\approx 3/\sqrt{2} \mu^{-1}\sqrt{N/3}$. This value is reported in panel (b) of Fig.~\ref{fig:3b112_beam_0_c1_P} and is seen to match remarkably well the position of the minimum.
 
Similarly to the standard two-beam case, for this configuration we see that the mean-field prediction of no evolution is recovered as $N\to\infty$ due to the divergence of the equilibration time scale $\tau$.
 
 In summary the system in Setup I displays the same ``freeze-out" behavior described in Ref.~\cite{Friedland:2006} for all angular distributions with $c\neq1$: the polarization vectors in each beam are only able to deviate from their initial values by a vanishing small amount in the large system size limit. The case with $c=1$ is peculiar in that we observe flavor conversion with a system size independent amplitude but a diverging time scale $\tau\propto\mu^{-1}\sqrt{N}$. As we will see in the next section, the presence of instabilities in the system from Setup II, for appropriate values of $c$, will change this picture qualitatively.

\subsection{Setup II} 
 \label{sec:psi2}
 
 For setup II the initial product state reads as
 \begin{equation}
 \ket{\Psi_{2}(0)} = \ket{\uparrow}^{\otimes N/3}\otimes\ket{\downarrow}^{\otimes N/3}\otimes\ket{\uparrow}^{\otimes N/3}\;.
 \end{equation}
The angular configurations with $c=-1$ is equivalent to the same angle in the previous setup (upon exchanging $B\Leftrightarrow C$) and large flavor conversion in beams $B$ and $C$ is seen with a typical time scale $\tau\propto\mu^{-1}\sqrt{N}$. The configuration with $c=1$ is instead equivalent to $c=0$ of the previous setup which, as discussed in the previous section,  behaves similarly to the other stable cases in Setup I with a decaying amplitude of flavor oscillations as a function of system size for all beams. In this case the time evolution is however twice as fast due to the presence of an additional factor of $2$ in the Hamiltonian (see Eq.~\eqref{eq:czero_setup1} and Eq.~\eqref{eq:c1_s1}).
 
 For angular distributions with $c\neq\pm1$ we can predict the qualitative behavior of the flavor evolution using the same line of reasoning used to obtain the effective Hamiltonian in Eq.~\eqref{eq:twob_ac} above. We first rewrite the Hamiltonian as a sum of two commuting parts to which we have added an unimportant constant, $h=5\mu\left(6+N\right)/6$,
 \begin{equation}
 \begin{split}
 \label{eq:3beam_as2}
H_{ABC} + h&= \frac{\mu}{N} \vec{J}^2+\frac{\mu}{N}\left(\vec{J}^2_{AB}-2c\vec{J}_C\cdot\left(\vec{J}_A-\vec{J}_B\right)\right)\\
&= \frac{\mu}{N} \vec{J}^2+\frac{\mu}{N}\left(\vec{J}^2_{AB}-2c\vec{J}\cdot\left(\vec{J}_A-\vec{J}_B\right)\right)\\
&:= \frac{\mu}{N} \vec{J}^2+H^{\rm dynamic}_{ABC}\;,
\end{split}
 \end{equation}
 where in the second line we used the fact that $\vec{J}_A^2$ and $\vec{J}_B^2$ are conserved and take the same value on our initial state.
 As already commented, the contribution proportional to the total angular momentum does not lead to oscillations in the large $N$ limit and all the flavor dynamics for $c\neq\pm1$ is driven by the second term denoted as $H^{\rm dynamic}_{ABC}$. This dominant part of the Hamiltonian is reminiscent of the two-beam Hamiltonian describing bipolar oscillations~\cite{Roggero:2021,PhysRevD.104.123023,martin2021classical}
\begin{equation}
H_{\rm bip} = \frac{2\mu}{N}\vec{J}^2_{AB}-\delta_\omega\vec{B}\cdot\left(\vec{J}_A-\vec{J}_B\right)\;,
\end{equation}
with the constant vector $\vec{B}$ replaced by the total spin of the system $\vec{J}$. In this expression, $\delta_\omega$ is proportional to the vacuum energy difference in the two beams. We can now show that for low energies and large system sizes, the Hamiltonian $H^{\rm dynamic}_{ABC}$ in Eq.~\eqref{eq:3beam_as2} has the same properties as $H_{\rm bip}$ and in particular shows the same phase transitions discussed in~\cite{PhysRevD.104.123023}.
Thanks to the all-to-all couplings in the Hamiltonian, the ground state can be approximated accurately with a mean-field state (see~\cite{Brandao2016} and~\cite{PhysRevD.104.123023}) so that its energy, can be written as
 \begin{equation}
 \begin{split}
E_0(c)&= \frac{2\mu}{N}\langle\vec{J}_A\rangle\cdot\langle\vec{J}_B\rangle-\frac{2c\mu}{N}\langle\vec{J}\rangle\cdot\left(\langle\vec{J}_A\rangle-\langle\vec{J}_B\rangle\right)\\
&=\frac{2\mu}{N}\langle\vec{J}_A\rangle\cdot\langle\vec{J}_B\rangle-\frac{c\mu}{3}\left(\langle J^z_A\rangle-J^z_B\rangle\right)\;,
 \end{split}
 \end{equation}
 where we used $\langle\vec{J}\rangle = (0,0,N/6)$. This is exactly the same energy function one obtains with $H_{\rm bip}$ and displays the same quantum phase transitions (see~\cite{PhysRevD.104.123023}).
 In particular, for the initial state $\ket{\Psi_2(0)}$, we expect to see a dynamical phase transition for $-1<c\leq0$ with substantial flavor oscillations and no dynamical flavor evolution for $0<c<1$. This is compatible with the expectations from the mean-field linear stability analysis discussed in Sec~\ref{sec:lin_stab} with the exception of $c=0$ which was considered stable instead.
 
 The point $c=0$ is the critical point and dynamics there is expected to happen on time-scales $t\approx\sqrt{N}$, as in the marginally stable configurations with $c=\pm1$ in the previous setup, while for $-1<c<0$ flavor evolution should happen on time scales $t\approx\log(N)$. Notably, the frequency of bipolar oscillations generated by the dynamical phase transition in $H_{\rm bip}$ are proportional to $\sqrt{\mu\delta_\omega}$ which is typically much smaller than $\mu$ close to the neutrino-sphere. In the multi-angle case studied here instead, the coupling constant in front of the one body term in $H^{\rm dynamic}_{ABC}$ is also proportional to $\mu$ and this gives rise to oscillations with frequency proportional to $\mu$ instead. This suggests that the mechanism behind both bipolar and fast oscillations is the same dynamical phase transition and the difference in time-scales is simply given by the difference in coupling constants.

In order to better illustrate the similarity between the dynamical phase transition in the two-beam case leading to bipolar oscillations and the unstable configurations in the present three-beam setup, we now present results for the Loschmidt echo. This is defined as the (squared) overlap between the evolved state $\ket{\Psi(t)}$ and the initial state 
as follows
\begin{equation}
    \mathcal L(t) = \left|\langle \Psi(0) | \Psi(t) \rangle \right|^2\;.
\end{equation}
As discussed more in detail in Refs.~\cite{Heyl2013,Heyl2018} (see also Ref.~\cite{PhysRevD.104.123023} for applications in neutrino physics) a Dynamical Phase Transition is signalled by non-analyticities of the Loschmidt echo as a function of time. For systems with degenerate initial state, as both the two-beam bipolar case for $\delta_\omega=0$ or the three-beam unstable case for $c=0$, a suitable generalization of this quantity is obtained as follows (see Refs.~\cite{Heyl2014,Zunkovic2018,PhysRevD.104.123023}) 
\begin{equation}
\label{eq:echoes}
    \mathcal L_k(t) = \left|\langle\Phi_k | \Psi(t) \rangle \right|^2\;.
\end{equation}
where $\ket{\Phi_k}$ are the two degenerate states: one is the initial state $\ket{\Phi_0}=\ket{\Psi(0)}$, and the other one is obtained by exchanging the polarization of the $A$ and $B$ beams. In our setup we have then $\ket{\Phi_1}=\ket{\downarrow}^{\otimes N_A}\otimes\ket{\uparrow}^{\otimes N_B}\otimes\ket{\uparrow}^{\otimes N_C}$ to $\ket{\Psi_2(0)}$. From these definitions of the Loschmidt echo we can also introduce a related quantity, the Loschmidt rate, defined as
\begin{equation}
\lambda(t) = -\frac{1}{N} \log[\mathcal L(t)]\;.
\end{equation}
Here $N$ is the total number of particles in the system and $\lambda(t)$ an intensive "free energy"~\cite{Heyl2013,Gambassi2012}. The rate $\lambda(t)$ plays here the role of a non-equilibrium equivalent of the thermodynamic free-energy. In cases where the generalization of the Loschmidt echo from Eq.~\eqref{eq:echoes} applies, the``free energy" is given by the minimum of the two rates $\lambda(t)=\min[\lambda_0(t),\lambda_1(t)]$ (see~\cite{Heyl2014}). In these cases, a dynamical phase transition can therefore occur whenever these rates cross for some time $t^*$, giving rise to a kink in $\lambda(t)$. \begin{figure}
    \centering
    \includegraphics[width=0.48\textwidth]{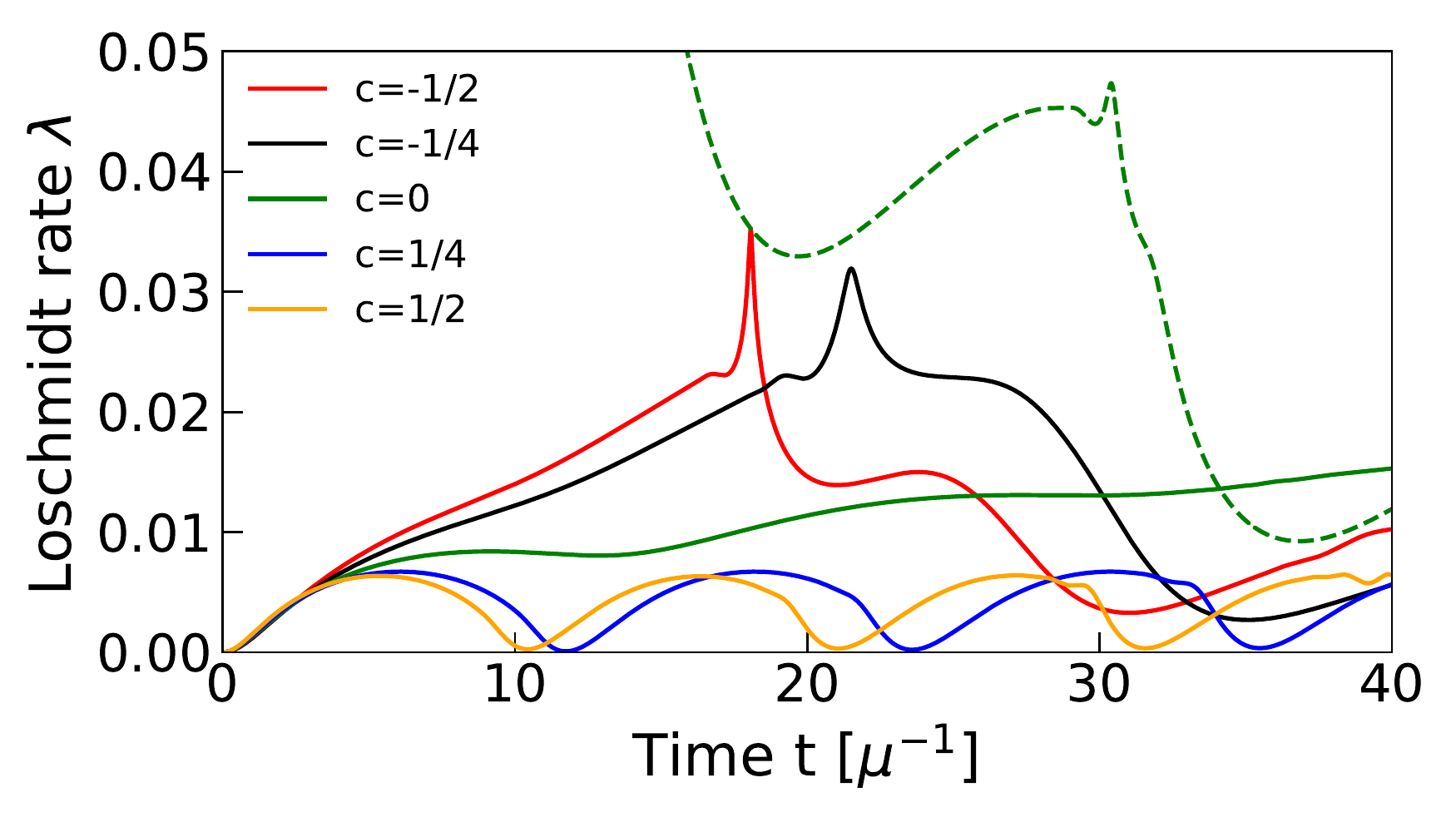}
    \caption{Loschmidt rate in a system with $N=348$ neutrino amplitudes initialized in $\ket{\Psi_2(0)}$ of setup II with different values of $c$: the red and black lines correspond to unstable cases with negative values $c=-1/2$ and $-1/4$ respectively, the blue and orange lines correspond to stable cases with positive values $c = 1/4$ and $1/2$ respectively, and the green line is with $c=0$ transiting from unstable to stable configuration.
    Solid lines are for the Loschmidt rate $\lambda(t)$ while dashed line shows the second rate $\lambda_1(t)$ as defined in the text.}
    \label{fig:loschmidt_rate}
\end{figure}
We present results for these Loschmidt rates at various values of the angular parameter $c$ in Fig.~\ref{fig:loschmidt_rate}. The second Loschmidt rate $\lambda_1(t)$ is shown only for the degenerate case $c=0$. These results can be directly compared with Fig.9 of Ref.~\cite{PhysRevD.104.123023} where the two-beam setup was considered instead. Similarly to that situation, we find that indeed the Loschmidt rates cross for a time $t^{*}\approx34\mu^{-1}$ for $c=0$ while for non-zero values of $c$ the behavior is markedly different between the stable and unstable cases: for stable configurations with $c>0$ the Loschmidt rate displays periodic oscillations that return to zero while for unstable configurations the Loschmidt rate shows sharp peaks. This is exactly the behaviour found in Ref.~\cite{PhysRevD.104.123023} for the case of slow bipolar modes and suggests that the argument provided above, which links this phenomenon to the fast oscillation case as being produced by the same dynamical phase transition, might be valid. Further work to establish a more robust connection and explore the full dynamical phase diagram of the model is warranted.

\begin{figure}
    \centering
    \includegraphics[width=0.48\textwidth]{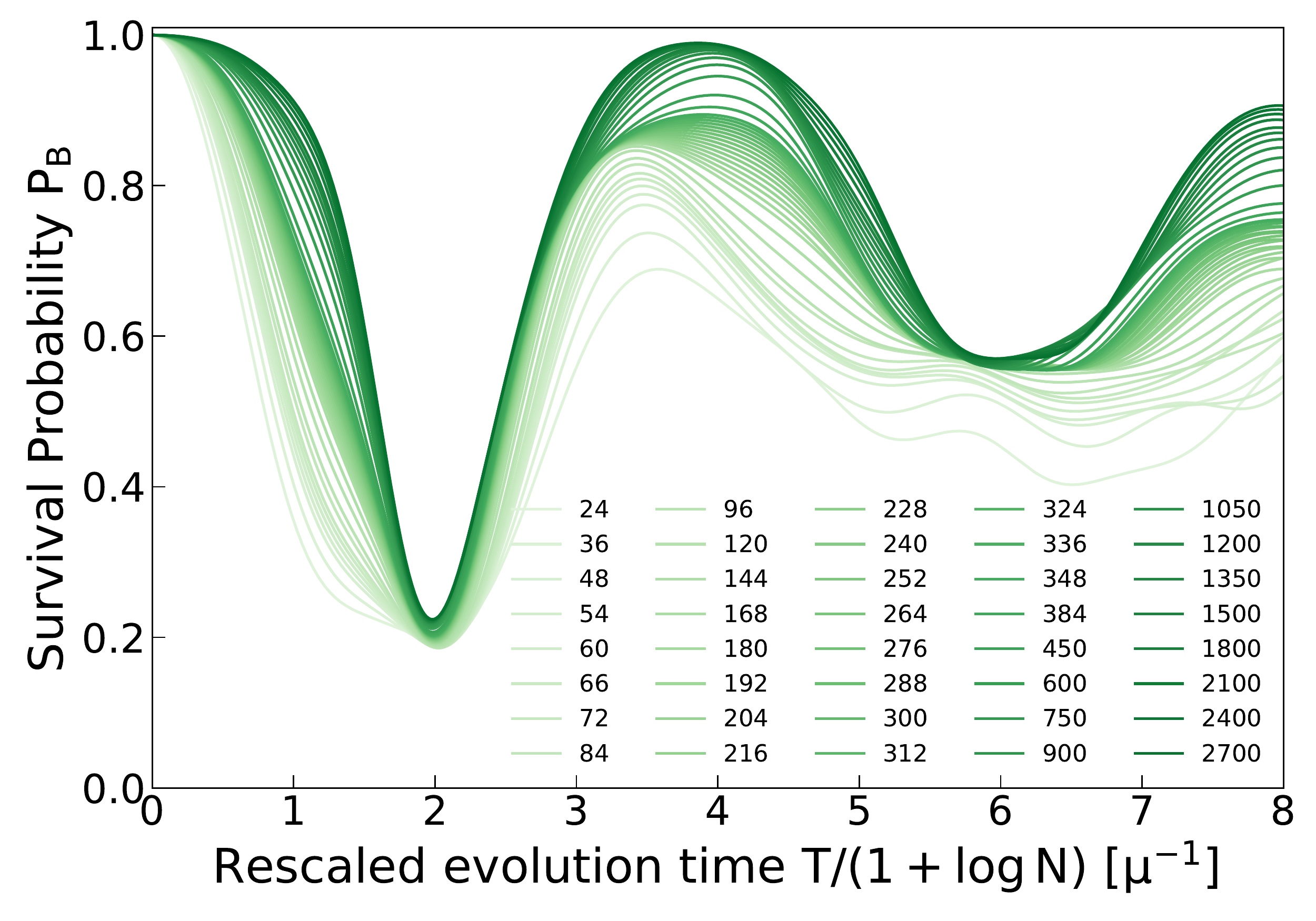}
    \caption{Time evolution of survival probability in beam $B$ for the initial state in Setup II and $c=-0.5$. The plot uses a rescaled time variable $T/(\rm 1+\log N)$ (darker colors indicate larger values of $N$) to highlight the system size dependence.}
    \label{fig:pb_121_m0p5_norm}
\end{figure}

\begin{figure}[b]
    \centering
    \includegraphics[width=0.48\textwidth]{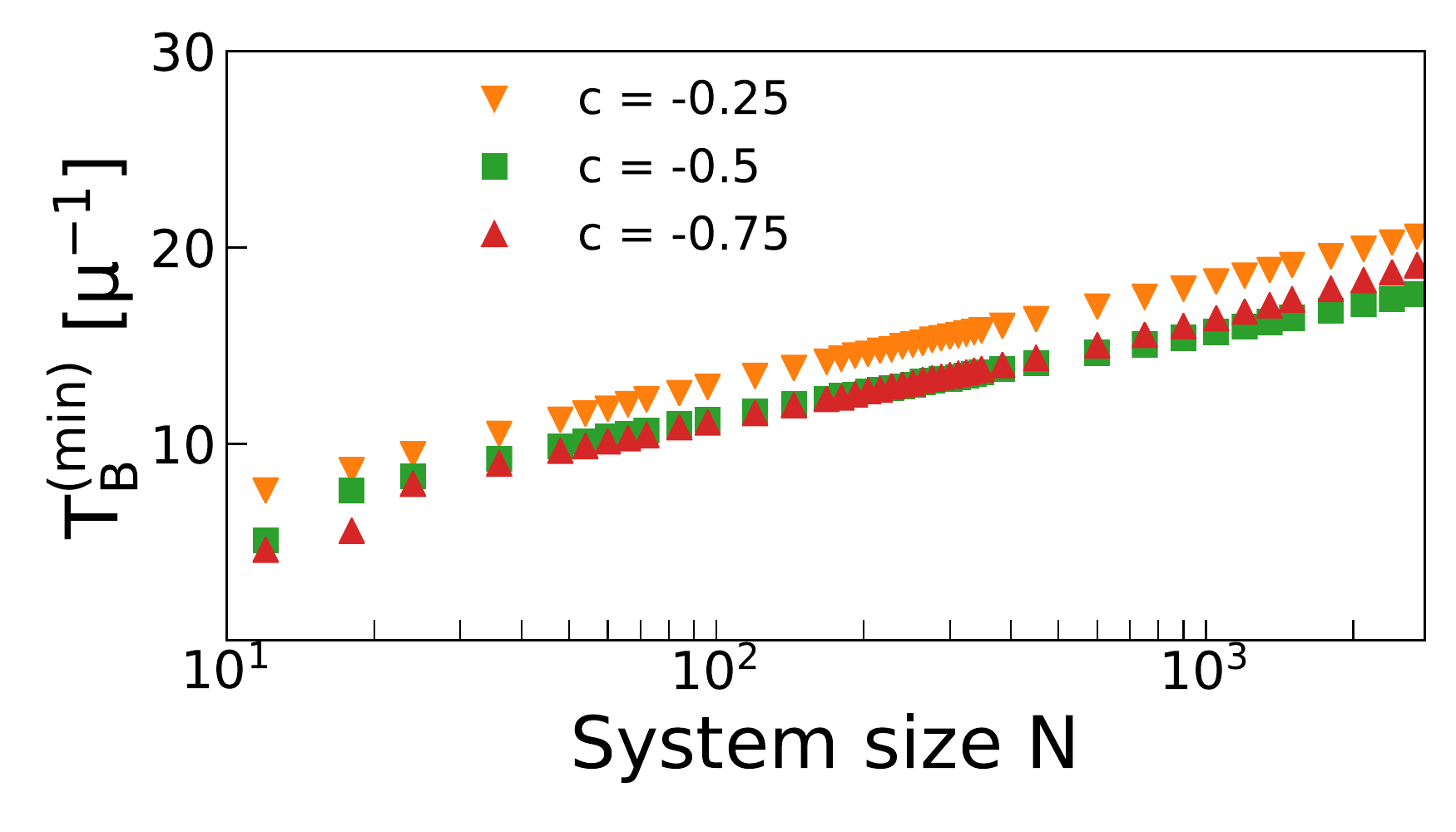}
    \caption{Time to reach the first minimum in the survival probability in beam $B$ for the initial state in Setup II and different angular distributions as function of system size (on a log scale). The straight lines for $-1<c<0$ emphasize the $\rm \log N$ dependence for unstable configurations.}
    \label{fig:time_pb_121_m0p5_norm}
\end{figure}
  \begin{figure}[t]
    \centering
    \includegraphics[scale=0.45]{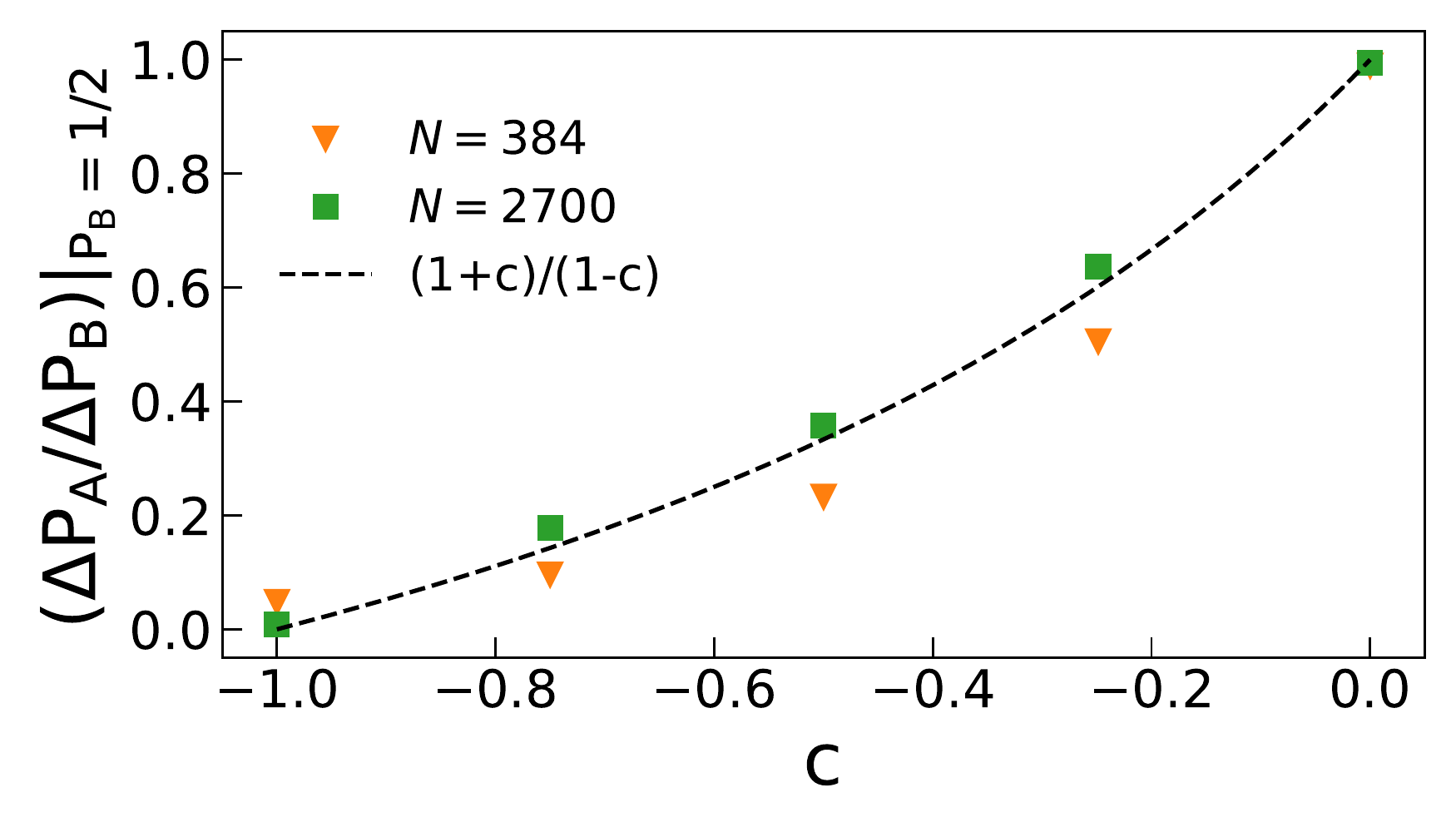}
    \caption{Setup II: Comparison between the relation of $|Q_A|^2/|Q_B|^2$ in mean-field approximation and the ratio of transition probabilities, $\Delta P_A/\Delta P_B$, in many-body calculations for five unstable parameters of $c$ and two system sizes $N=384$ and 2700, respectively.}
    \label{fig:121_beams_survival_probability}
\end{figure}

 \begin{figure*}[t]
 \subfloat[$c=1$]{\includegraphics[width=0.33\textwidth]{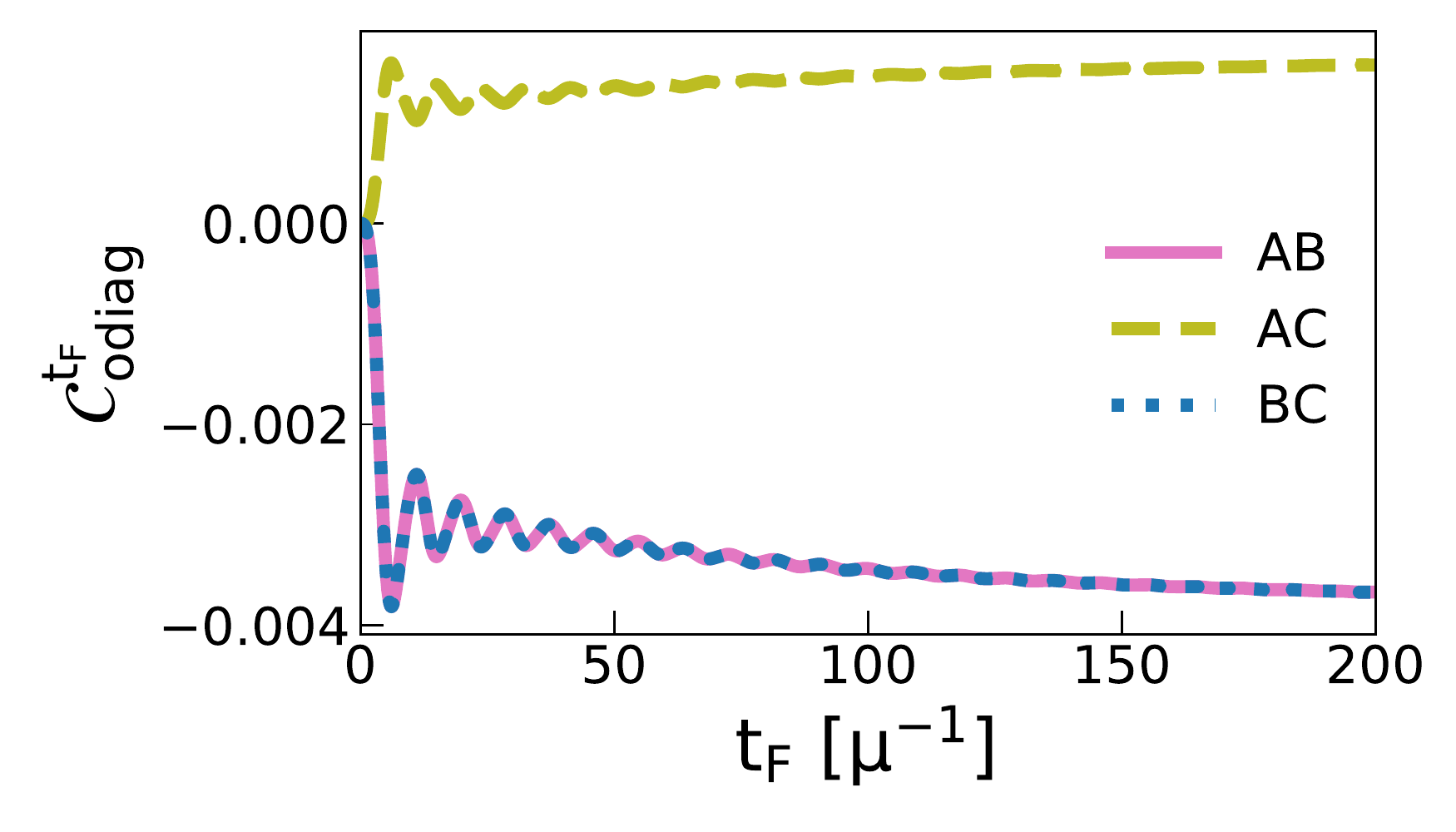}}
 \subfloat[$c=-1$]{\includegraphics[width=0.33\textwidth]{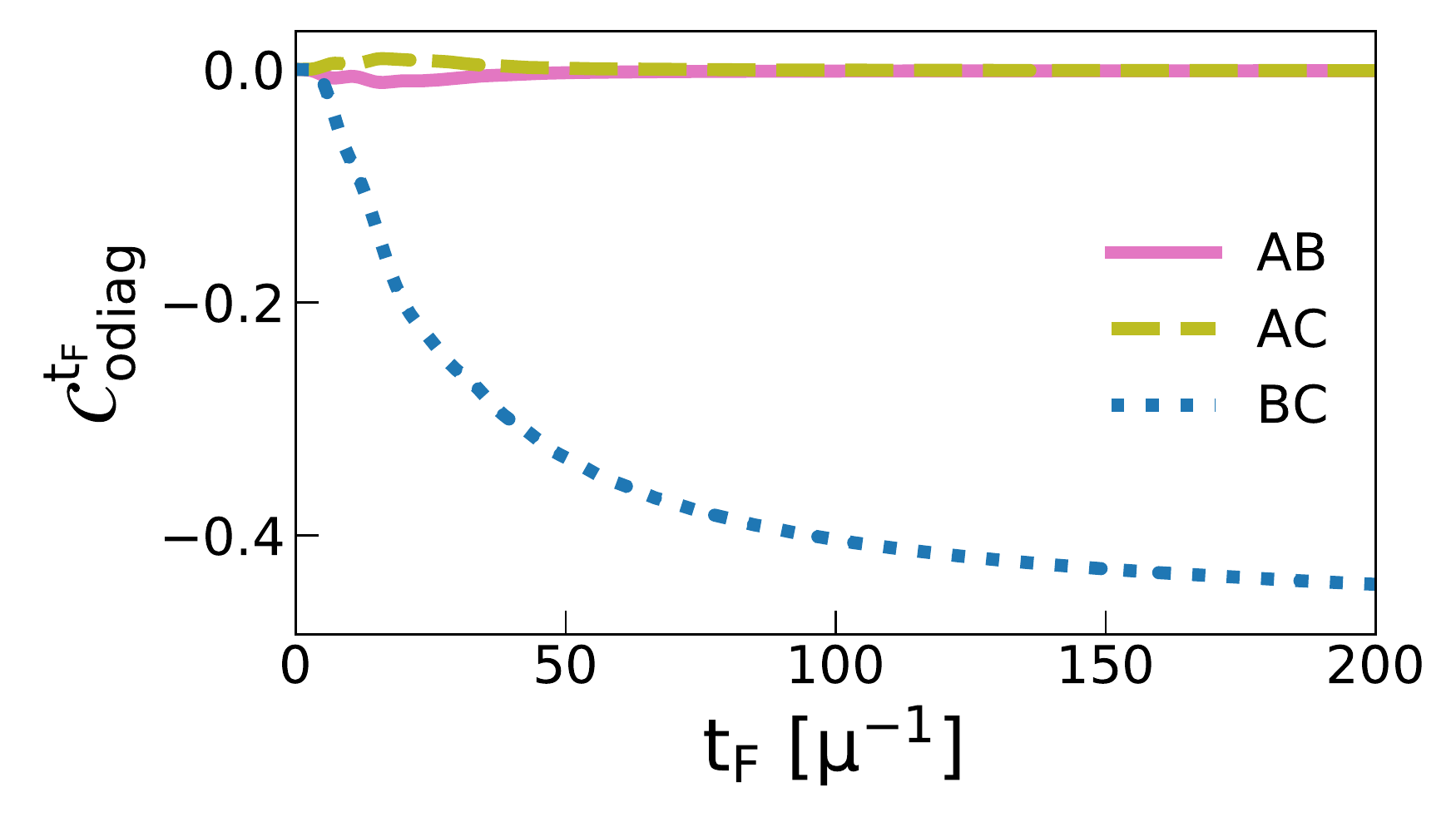}}
 \subfloat[$c=-0.5$]{\includegraphics[width=0.33\textwidth]{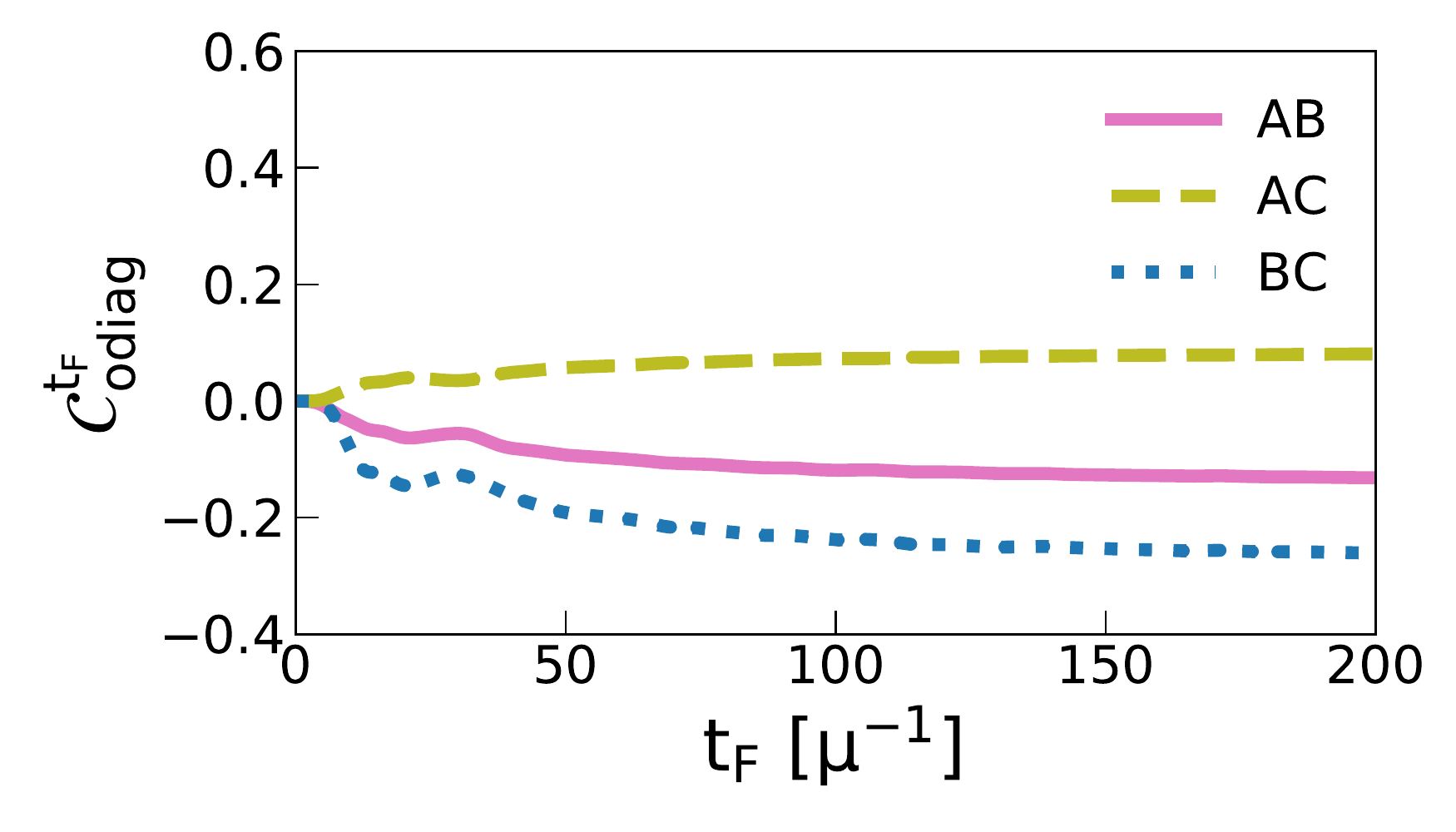}}
 \caption{The time evolution of the three off diagonal pair correlations $\mathcal{C}_{AB}$, $\mathcal{C}_{AC}$ and $\mathcal{C}_{BC}$ (shown with full, dashed and dotted lines respectively) for the system size $N=348$. The left panel is a stable system with $c=1$; the center panel depicts a marginally unstable case with $c=-1.0$, and the right panel shows an unstable case with $c=-0.5$.}
 \label{fig:odiag_corrs}
\end{figure*}

As a further confirmation of the stability of configurations with positive values of $c$, the numerical results we obtain for these configurations show indeed flavor evolution similar to the stable cases observed before, with oscillation amplitudes vanishing as $\approx1/N$ in the large system size limit.

 We can now turn to present the result of our simulation for the survival probability in beam $B$ and $c=-0.5$ in Fig.~\ref{fig:pb_121_m0p5_norm}. The time axis has been scaled by the factor $\rm 1+\log N$ to show the system size dependence (darker curves corresponds to larger systems).  
 In Fig.~\ref{fig:time_pb_121_m0p5_norm} we plot the system size dependence of the time to reach the first minimum in the survival probability in beam $B$.  As expected from the discussion on the presence of a dynamical phase transition in this regime, the time dependence for the unstable configurations with $-1<c<0$ is logarithmic $T^{\rm min}_B\propto\log(N)$. We have observed the same qualitative behavior for all beams and values of $c$ in the unstable region. 

The ratios of flavor transitions in each beams during that intermediate stage and their dependence on the angular parameter $c$ are also compared to the mean-field relations in Eq.~\ref{eq:linear_analysis_eigenvector}.
We pick a time point where the transition probability of beam $B$, $\Delta P_B \equiv |1-P_B| $, firstly reaches a value of $1/2$ ($\sim \mathcal O(1)$) to represents an intermediate stage.
We calculate the ratio of transition probabilities, $\Delta P_A/\Delta P_B$, at that time, and do the same for all five unstable angular parameters of $c\leq 0$ and two system sizes $N=384$ and 2700 in Fig.~\ref{fig:121_beams_survival_probability}.
At a larger system size of $N=2700$, the ratios tend to converge on the prediction from the linear analysis.

However, the long-term evolution of survival probability can deviate from the mean-field result.
Within mean-field assumption, this three-beam setup where two beams are anti-aligned is equivalently an axisymmetric setup and should lead to a bipolar motion with the same minimum survival probabilities in each flavor conversion cycle \cite{Johns:2020,padilla2022neutrino}, but Fig.~\ref{fig:pb_121_m0p5_norm} shows that the minimal value of survival probabilities in the second cycle (at the rescaled time $\approx 6$) is much higher than that in the first one (at the rescaled time $\approx 2$) as the system size $N$ goes to 2700.
This deviation is consistent with the behavior of decoherence found in Ref.~\cite{Xiong:2021} and will also be reflected by the entanglement and correlations as to be discussed in next section.

\section{Entanglement and correlations}
\label{sec:entanglement_correlations}
In the previous section we studied the dependence of single particle observables like the survival probability on system size. For marginally unstable and unstable configurations we discovered that the many-body result does not converge to the mean field prediction in the large particle number limit. 
When such a difference appears, one is left to wonder whether the initial mean field wavefunction evolves with time to a more complicated one. In such a scenario, many-body effects like correlations and entanglement, which would otherwise not be present, tend to develop dynamically~\cite{Rrapaj:2020,Roggero:2021,PhysRevD.104.123023,PhysRevD.104.123035}. The focus of this section is the study of the pair correlations and entanglement generated during the time evolution.

\subsection{Beam Correlations}

As we mentioned in the derivation of the mean-field equations in Sec.~\ref{sec:lin_stab}, one of the underlying assumptions behind the mean field approximation is the factorization of expectation values $\langle \mathcal{O}_i\mathcal{O}_j\rangle\approx\langle \mathcal{O}_i\rangle\langle\mathcal{O}_j\rangle$ for different neutrinos. Here we explore the violations of this assumptions due to many-body effects by measuring the connected pair correlations along the flavor axis
\begin{equation}
\mathcal{C}_{A_iA_j} = \frac{4}{N_{A_i}N_{A_j}}\left(\langle J^z_{A_i}J^z_{A_j}\rangle- \langle J^z_{A_i}\rangle\langle J^z_{A_j}\rangle\right)\;.
\end{equation}
We first note that, due to the conservation of the total polarization along the z-axis, the sum for all $A_i,A_j\in\{A,B,C\}$ becomes
\begin{equation}
\sum_{A_i,A_j} \mathcal{C}_{A_iA_j}(t) = \frac{4}{N^2}\left(\langle (J^z)^2\rangle - \langle J^z\rangle^2\right)=0\;, 
\end{equation}
where the last equality comes from the initial condition being a product state. This constraint implies that the intra-beam correlations $\mathcal{C}_{A_iA_i}(t)$ are not independent on the correlations $\mathcal{C}_{A_iA_j}(t)$ between different beams $A_i\neq A_j$. In particular we have
\begin{equation}
\label{eq:cdiag_odiag}
\mathcal{C}_{diag}(t) = -\mathcal{C}_{odiag}(t);,
\end{equation}
where $\mathcal{C}_{diag}(t)$ and $\mathcal{C}_{odiag}(t)$ are the sum of diagonal and off-diagonal pair correlations respectively.

As we have seen in the previous section, for appropriate values of the angular parameter $c$ the three-beam models considered in this work can show flavor evolution in contrast to the mean-field prediction. In these situations we, then, expect correlations to be present as they are responsible for the non-trivial evolution. Since the system in Setup II can reproduce all three types of time evolution (stable, marginally unstable and unstable) we restrict the present discussion to this setup only.

  \begin{figure}[t]
 \includegraphics[width=0.48\textwidth]{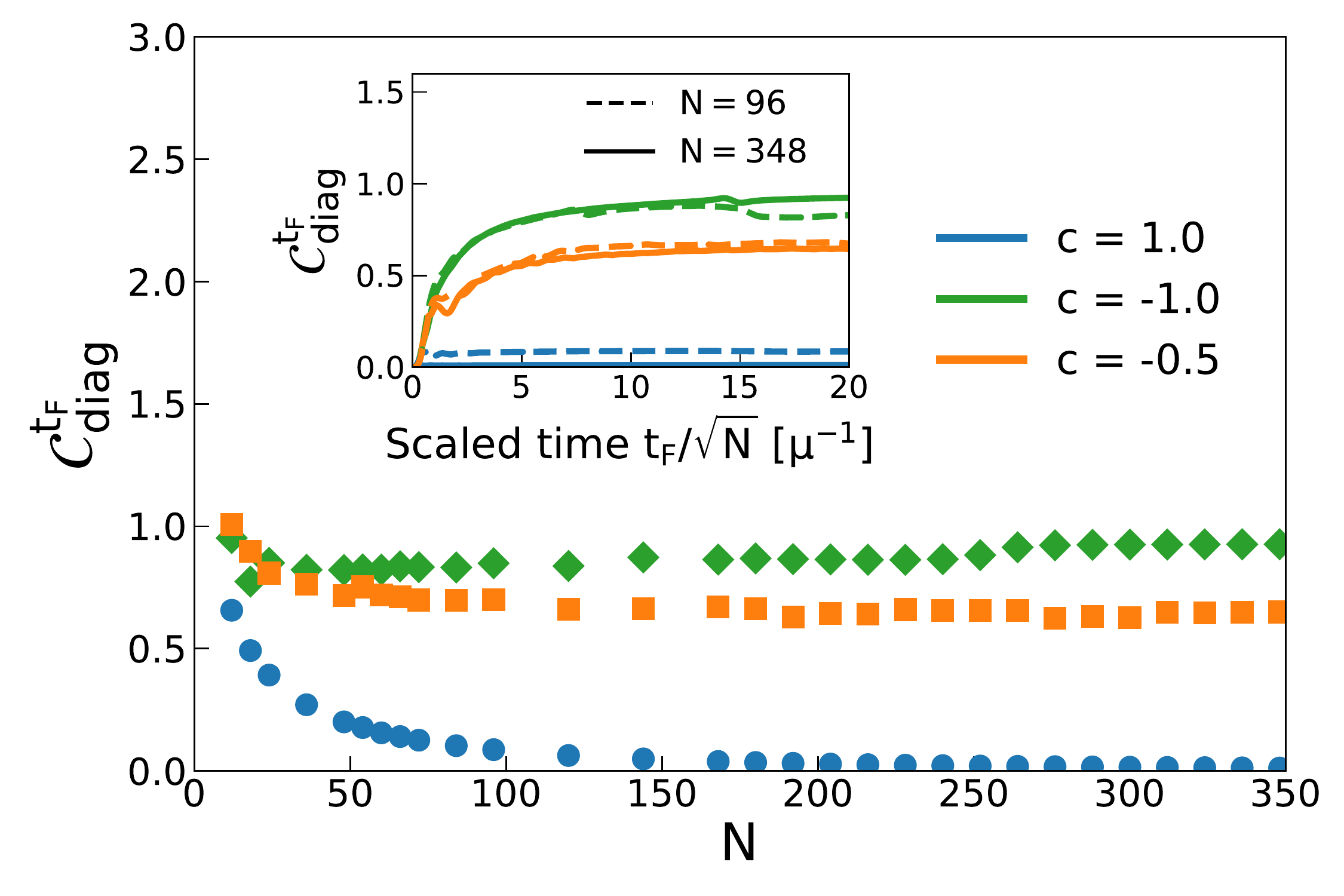}
\caption{Evolution with system size $N$ of the long time averaged diagonal pair correlation $\mathcal{C}^{t_F}_{diag}$ from Eq.~\eqref{eq:diag_c_time_avg} for the initial state $\ket{\Psi_2(0)}$ and three angular distributions: $c=1$ (blue circles), $c=-1$ (green diamonds) and $c=-0.5$ (orange squares). The inset shows the dependence on the integration time $t_F$ for the three angular distributions and two system sizes: $N=96$ (dashed lines) and $N=348$ (solid lines). The time axis in the inset has been scaled with $1/\sqrt{N}$ to better show the systems size dependence.}
 \label{fig:sum_aiag_avg}
 \end{figure}

In Fig.~\ref{fig:odiag_corrs} we show the time evolution of all three off-diagonal pair correlations for three indicative scenarios and $N=348$: the left panel shows the case of a stable system with vanishing flavor evolution ($c=1$), the central panel shows results for a marginally unstable system with flavor evolution at the long time scale $\tau\approx\mu^{-1}\sqrt{N}$ ($c=-1$) and the right panel shows results for an unstable system ($c=-0.5$) with flavor evolution at the short scale $\tau\approx\mu^{-1}\log(N)$. These results shows that for stable systems $C_{A_iA_j}\approx 0$ at all times, with like flavor beams ($A$ and $C$) positively correlated and opposite flavor beams anti-correlated. For the marginally unstable system at $c=-1$ the two anti-parallel beams $B$ and $C$ whose total spin is conserved are strongly anticorrelated and along times $C_{BC}\approx -0.5$ while the stable beam $A$ has vanishing correlation with the other two. Finally, for the unstable case $c=-0.5$, all beams show substantial correlations among each other.

The results shown in Fig.~\ref{fig:odiag_corrs} suggest that one can detect instabilities in the neutrino flavor evolution by looking at pair correlations among the beams while the conservation of the total spin also indicates (see Eq.~\eqref{eq:cdiag_odiag}) that correlations must be present inside the beams themselves. These correlations are however influenced by finite size effects and for small system sizes this separation is less pronounced. To show this we present in Fig.~\ref{fig:sum_aiag_avg} the long time average of the total diagonal correlations
\begin{equation}
\label{eq:diag_c_time_avg}
\mathcal{C}^{t_F}_{diag} = \frac{1}{t_F} \sum_{A_i}\int_0^{t_F}dt \mathcal{C}_{A_iA_i}(t)\;,
\end{equation}
as a function of system size $N$. Due to Eq.~\eqref{eq:cdiag_odiag} we have that $\mathcal{C}^{t_F}_{diag}=-\mathcal{C}^{t_F}_{odiag}$ and the quantities provide a similar measure of correlations. The main panel shows $\mathcal{C}^{t_F}_{diag}$ in the three cases considered above for $t_F=400\mu^{-1}$. This value was chosen to guarantee convergence to the long time average for the largest system considered here, $N=348$. In general we observe that convergent results can be obtained for all angular distributions choosing $t_F\propto\sqrt{N}$, this is shown in the inset of Fig.~\ref{fig:sum_aiag_avg} where we present the dependence of the time averaged correlations with the size $t_F$ of the time window upon rescaling with $\sqrt{N}$: the dashed lines correspond to $N=96$ and the continuous lines to $N=348$.

\subsection{Entanglement entropy}

Another important way to characterize correlations in a many-body system is to estimate the amount of entanglement generated during time evolution. From a practical point of view, entanglement controls the computational cost of classical tensor network methods to simulate the flavor dynamics of a neutrino system. An important example, already used in the study of collective neutrino oscillations in Refs.~\cite{Roggero:2021,PhysRevD.104.123023}, and more recently in~\cite{Cervia:2022}, are Matrix Product State (MPS) which can approximate efficiently (i.e. in polynomial cost) quantum states for which the R\'{e}nyi entropies $R_\alpha$ for any bipartition of the system grow at most logarithmically in the size of the bipartition~\cite{Schuch2008}. We will comment more on the efficiency of a MPS simulation of collective neutrino systems in the conclusions.

Quantum correlations like entanglement are more generally useful tools to analyze the structure of many-body neutrino systems and have been shown to be helpful in detecting the presence of bipolar collective modes in the past~\cite{Roggero:2021,PhysRevD.104.123023,martin2021classical}. These calculations were performed using only two beams and therefore only display slow modes. Here we are interested in extending this connection to fast modes instead, and therefore, might be important near the surface of a proto-neutron star where $\mu\gg\omega$ \cite{Sawyer:2005,Izaguirre:2017}.  In~\cite{PhysRevD.104.123035}, the authors found that the largest values of entanglement entropies occur for neutrinos with energies closest to the spectral split energy. 
\begin{figure}[b]
     \includegraphics[scale=0.4]{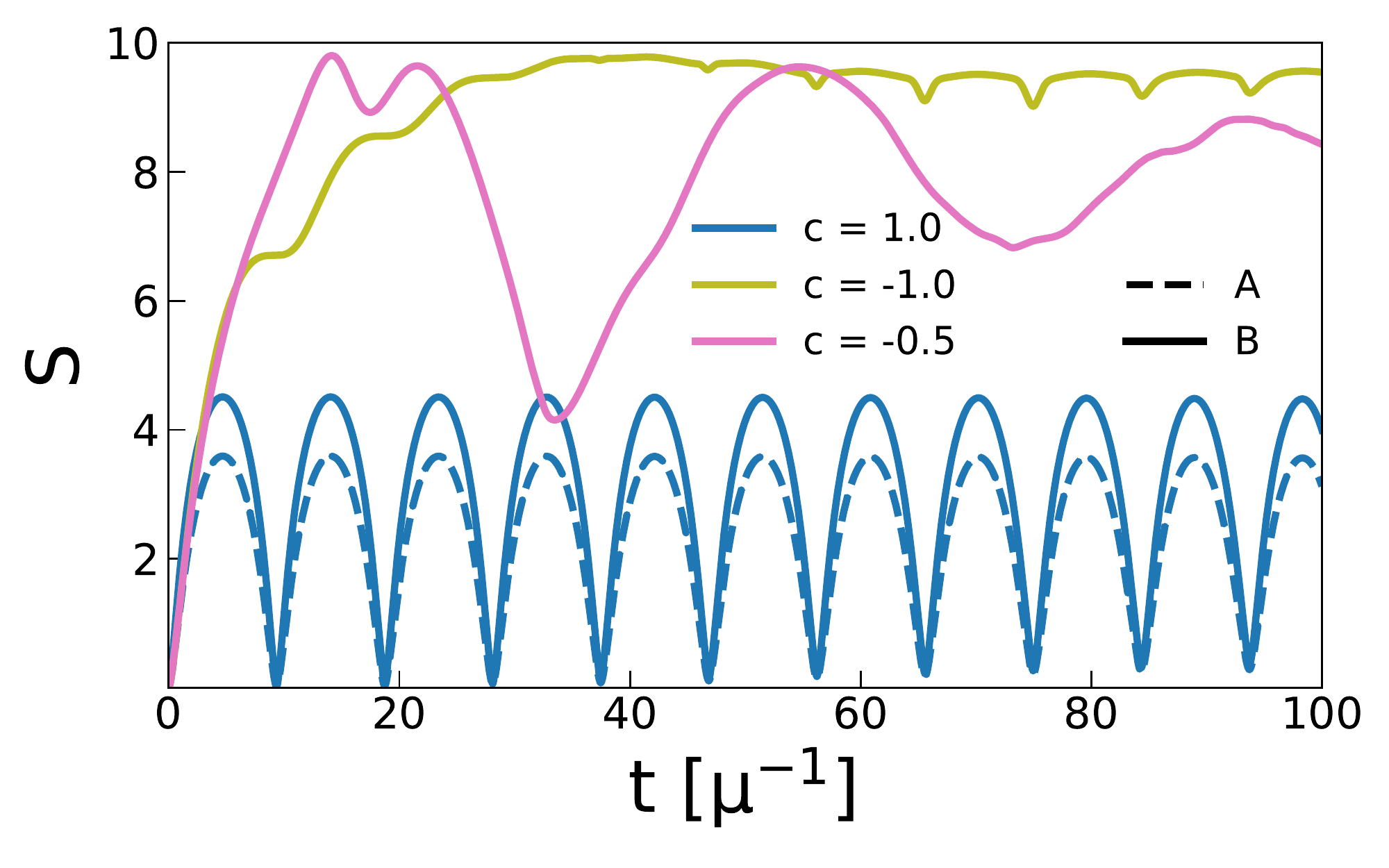}
     \caption{Entanglement entropy as function of time for system size $N=2700$. The three angular distributions are $c=1$ (blue), $c=-1$ (yellow) and $c=-0.5$ (purple).  Beam $A$ (dashed) is shown only for the  stable configuration ($c=1$) while beam $B$ is shown of all three angles.}
     \label{fig:r1_121_ba}
    \end{figure}
\begin{figure*}[ht]
 \subfloat[$c=1$]{\includegraphics[width=0.33\textwidth]{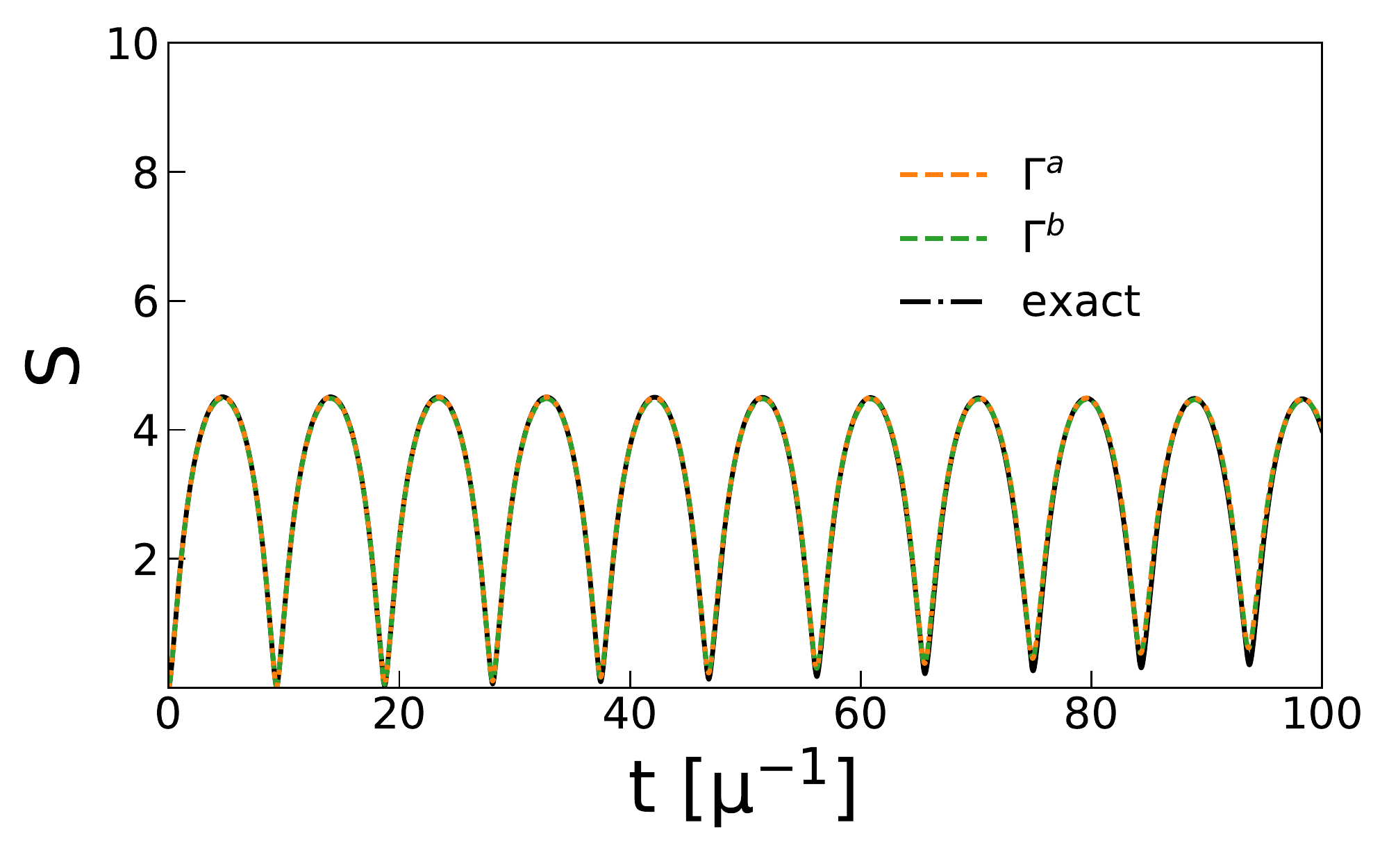}}
 \subfloat[$c=-1$]{\includegraphics[width=0.33\textwidth]{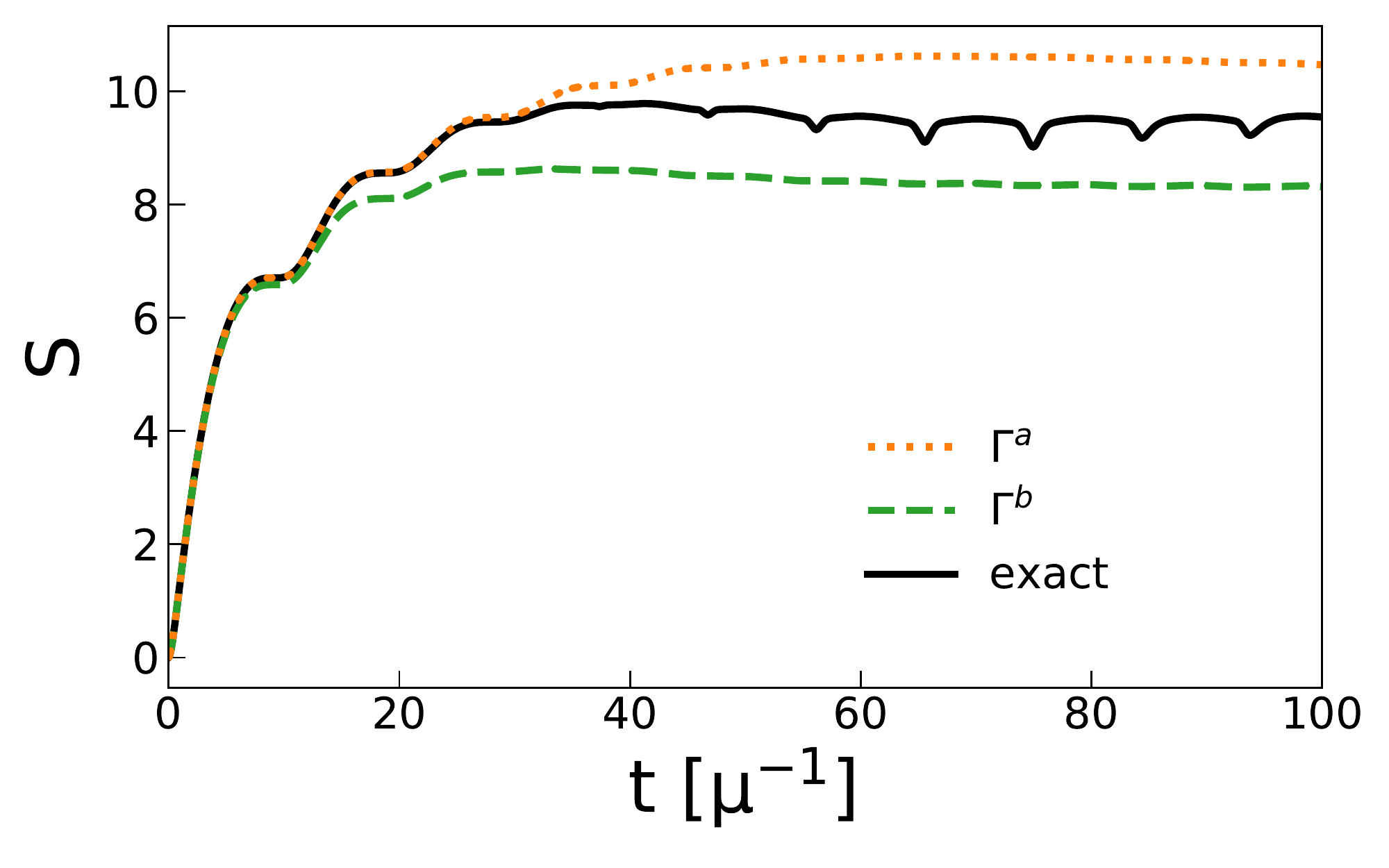}}
 \subfloat[$c=-0.5$]{\includegraphics[width=0.33\textwidth]{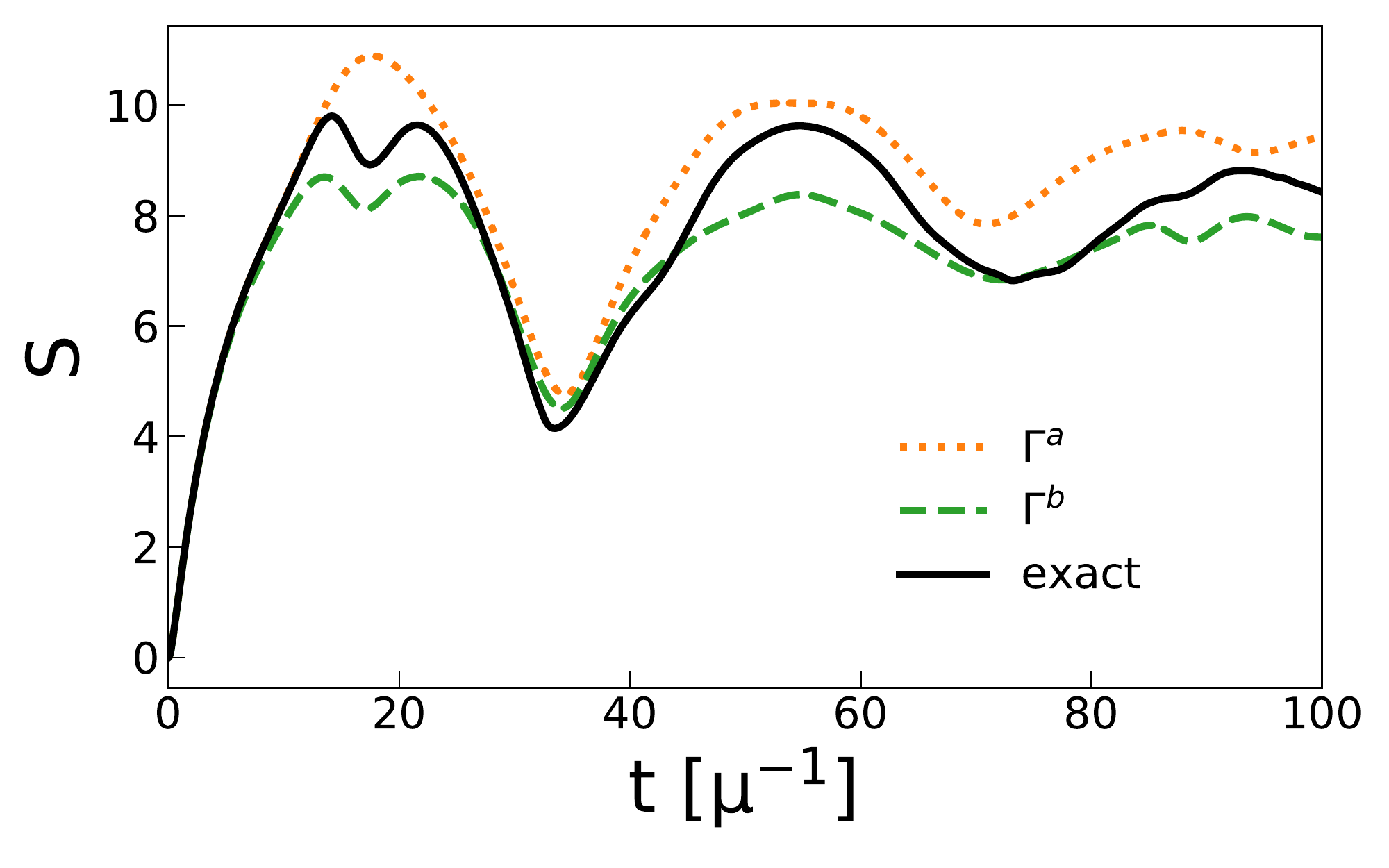}}
 \caption{The time evolution of the entanglement entropy for beam B and the system size $N=2700$. The left panel is a stable system with $c=1$; the center panel depicts a marginally unstable case with $c=-1.0$, and the right panel shows an unstable case with $c=-0.5$. The approximation $\Gamma^{(a)}_i(t) = N_i\left(1-P_i(t)\right)$ gives a closer result to the exact value than $\Gamma^{(b)}_i(t) = \frac{N_i}{4}-\frac{\langle J_i^zJ_i^z\rangle}{N_i}$.}
 \label{fig:R1_approx}
\end{figure*}

In Fig.~\ref{fig:r1_121_ba} we depict, as function of time for $N=2700$ and the three angular setups from the previous plots, the entanglement entropy $S_{A_i}$ (see eq.~\ref{eq:von_neumann_entropy_gen}) obtained from the reduced density matrix of the $A$ beam (full lines) and, for the stable system with $c=1$, also of the $B$ beam (dashed line). For this latter setup, in Figs.~\ref{fig:odiag_corrs} and~\ref{fig:sum_aiag_avg} we saw that correlations vanish as the system size increases while entanglement entropy does not. Instead, it rises quickly and  then proceeds to oscillate with a relatively small amplitude. For marginally unstable ($c=-1$) and unstable ($c=-0.5$) configurations, the entropy reaches $S^{max}\approx log_2(N/3)$. The associated timescales are  $t \sim \sqrt{N}$ and $t \sim \log(N)$ respectively, in agreement with our previous observations on the persistence in section~\ref{sec:psi2}. 

\begin{figure}[b]
     \includegraphics[scale=0.45]{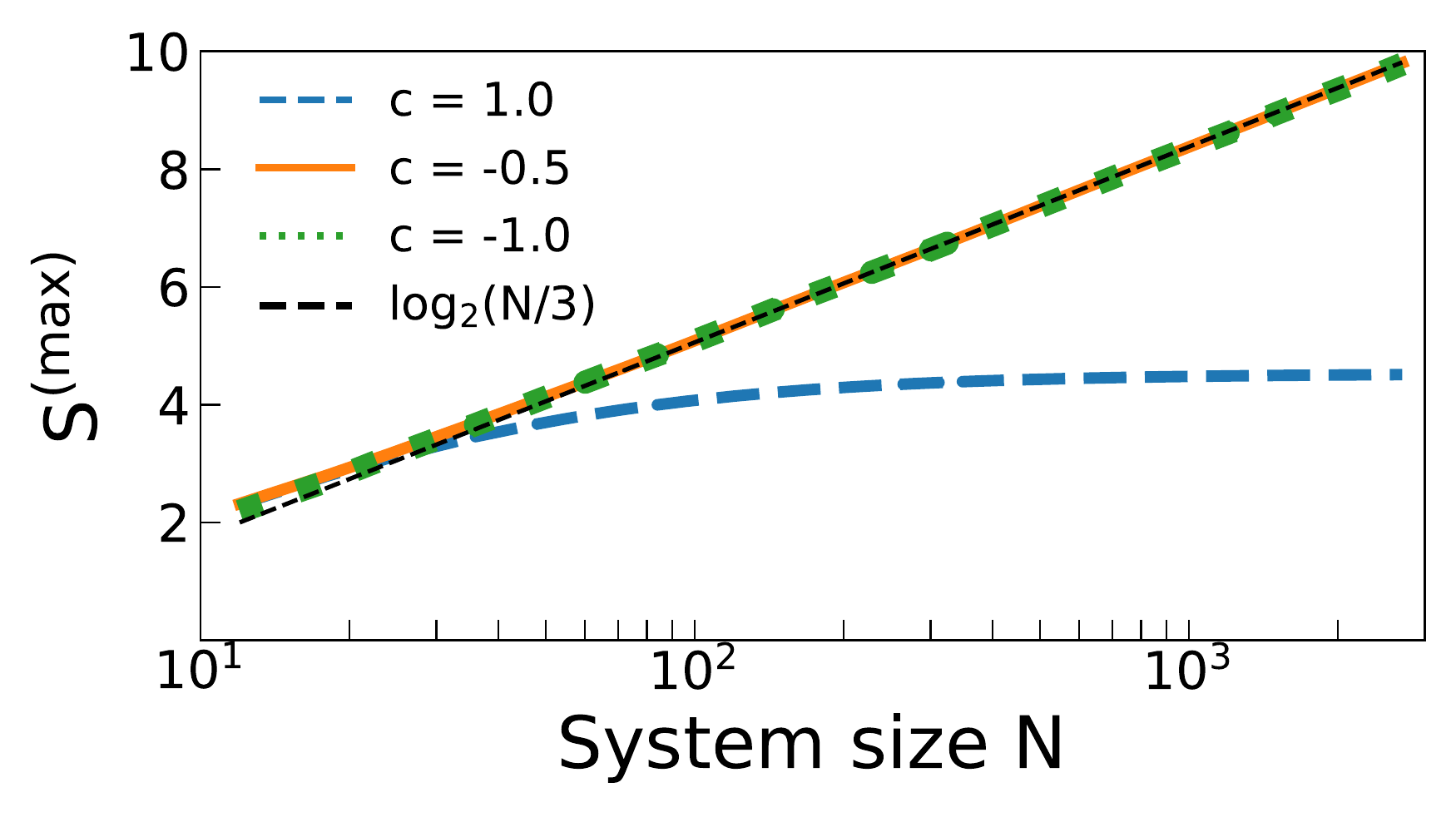}
     \caption{Maximal value of the entanglement entropy as function of $N$ for beam $B$, and angles $c=-1$ (dashed blue), $c=-0.5$ (solid orange), and $c=1$ (dotted green). For comparison, we have also included the $\log_2(N/3)$ (dashed black) functional form to show the system size dependence.}
     \label{fig:r1_max_121_ab_n}
    \end{figure}

To further confirm the behavior of $S^{max}$ with $N$, in Fig.~\ref{fig:r1_max_121_ab_n} we depict how it scales with system size, with x-axis in $\log$ scale,  for beam $B$. The stable configuration ($c=1$) reaches a plateau while the marginally unstable and unstable configurations increase logarithmically. This is even more evident by comparing the data from simulations to the line $\rm \log_2(N/3)$ (dashed black line). This logarithmic behavior has also been observed in past MPS based calculations of bipolar oscillations~\cite{Roggero:2021,PhysRevD.104.123023} as well as more general two-beam models~\cite{martin2021classical}.

As the particle number increases, so do the initial expectation values of $J_z$ and $\vec{J}^2$. Then, for large $N$ we expect to represent the flavor operators $J^{x,y,z}_{A_i}$ through canonical bosonic operators, following the Holstein-Primakoff transformation~\cite{Holstein:1940} truncated to leading order. If we approximate the state of each beam by a Gaussian, then, the entanglement entropy (Von Neumann entropy) for a beam can be approximated,
\begin{equation}
\begin{split}
S_{{A_i}}(t) &= \frac{1+2\Gamma_{A_i}(t)}{\log(2)}\text{arccoth}\left(1+2\Gamma_{A_i}(t)\right)\\
&+\frac{1}{2}\left[\log_2(\Gamma_{A_i}(t))+\log_2(1+\Gamma_{A_i}(t))\right]\;.
\label{eq:SAi_gaussian}
\end{split}
\end{equation}
The term $\Gamma_{A_i}$, related to the covariance matrix of the Gaussian state, can be approximated in two different ways,
\begin{equation}
 \begin{split}
  \Gamma^{(a)}_{A_i}(t) =& N_{A_i}\left(1-P_{A_i}(t)\right),\\
  \Gamma^{(b)}_{A_i}(t) =& \frac{N_{A_i}}{4}-\frac{\langle (J_{A_i}^z)^2\rangle}{N_{A_i}}\;,
 \end{split}
\end{equation}
based on the survival probabilities and correlations, respectively. The detailed analysis can be found in Appendix~\ref{app:hp}. 
In Fig.~\ref{fig:R1_approx} we plot the entanglement entropy for beam $B$ and $N=2700$ as function of time, together with the predictions obtained from these two approximations. The panel on the left is a stable system with $c=1$; the center panel depicts a marginally unstable case with $c=-1.0$, and the right panel shows an unstable case with $c=-0.5$. The survival probability and correlations employed in making this plot were computed with the method described in section~\ref{sec:methods}. 
 For stable configurations, Eq.~\eqref{eq:SAi_gaussian} seems to match the exact result for long term dynamics, while for marginally unstable and unstable cases it is a rather good approximation for short time dynamics. The approximation based on the survival probability seems to perform better than the one based on correlations in the marginally unstable case (panel (b)) as it follows the exact evolution for longer times. In the unstable case however, this approximation misses the double peak structure around $t=20\mu^{-1}$ associated with flavor inversion in the beam (ie. $P_A<0.5$), a situation that cannot be described at leading order in the semi-classical expansion (see Appendix~\ref{app:hp} for more details).
  
 These results show that non-trivial evolution of the survival probability is intimately connected to the presence of entanglement and correlations. It would be interesting to extend this approach to more complicated models including a non-diagonal one body Hamiltonian (vacuum frequency) as well as performing the full evolution within the semi-classical approach. In addition, it remains to be seen whether the Holstein-Primakoff approximation is the best choice to represent the system, and, perhaps the truncated Wigner approximation (see~\cite{POLKOVNIKOV:2010} and references therein) could also be employed. While their numerical implementation is beyond the scope of this work, if semi-classical treatments are good approximations in the large particle number limit, they may be of practical importance for simulations relevant for supernovae and neutron star binary mergers.

\section{Summary and Conclusion}
\label{sec:summary}
    
    In compact objects, core-collapse supernovae and neutron star mergers, neutrinos play a vital role in shaping the dynamics of the system and the conditions for nucleosynthesis in the ejected material. The presence of collective flavor oscillations, primarily due to neutrino-neutrino scattering, could lead to important effects in these aforementioned developments. As these scatterings are dependent only on the angle between particles and couple neutrinos of different energies, flavor
evolution is a very complicated  many-body problem. 

In this work we performed a complete many-body treatment based on the method described in Ref.~\cite{Xiong:2021}.  We considered a simplified setup of three coplanar beams, parameterized by $c$ -- the cosine of the angle between two of the beams, and two neutrino flavors. We focused only the effects due to neutrino scattering and studied the dependence on system size.  We selected two initial configurations for the wavefunction, 
 \begin{enumerate}
  \item $\ket{\Psi_{1}(0)} = \ket{\uparrow}^{\otimes N_A}\otimes\ket{\uparrow}^{\otimes N_B}\otimes\ket{\downarrow}^{\otimes N_C}$,
  \item $\ket{\Psi_{2}(0)} = \ket{\uparrow}^{\otimes N_A}\otimes\ket{\downarrow}^{\otimes N_B}\otimes\ket{\uparrow}^{\otimes N_C}$.
 \end{enumerate}
 For these setups, the mean field approximation predicts no flavor evolution, and any dynamics is purely a many-body effect.
 By analyzing the time evolution of the survival probability or persistence for each of the beams, we discovered that for $|c|<1$ in setup I and $c \geq 0$ in setup II, the many-body results converge to the mean field ones in the large particle number limit. The extremal values of the cosine in setup I and $c=-1$ in setup II denote marginally unstable situations where the survival probability does not converge to $1$ as would have been predicted by the mean-field analysis. However, the time to reach its minimum $\sim \mu^{-1} \sqrt{N}$, ``freezing'' the flavor evolution for large $N$. The effective Hamiltonian governing these situations is analogous to the two beam system studied in Refs.~\cite{Roggero:2021,PhysRevD.104.123023} and the results agree with the behaviour found there.
 The system in setup II can instead develop fast collective oscillations when $-1<c<0$ leading to a crossing of the angular distributions. For these cases the persistence does not converge to $1$ as $N$ increases and the time scale to reach the first minimum $\sim \mu^{-1} \log(N)$. These unstable configurations can also be derived by rewriting the Hamiltonian to separate the dominant term, which is analogous to bipolar oscillations in the presence of the vacuum term. This establishes a connection between the dynamical phase transition leading to bipolar oscillations and the presence of fast modes. The main difference between the two situations is that now the one body term playing the role of the vacuum frequency there has a coupling proportional to $\mu$. This fact explains in a natural way the distinction between the frequency of oscillations in slow and fast modes.
 These results are also in agreement with the linear mean field instability analysis we performed in section~\ref{sec:lin_stab}. In a follow up work we plan to perform a detailed study of the dynamical phase diagram in this simple multi-beam model.
 
 To further confirm the presence of many-body effects, and beyond mean field behavior, we analyzed the entanglement entropy of each beam as well as the time averaged flavor correlations among them.
 The pair correlations agree qualitatively with the results of the persistence analysis: with increasing system size they vanish for stable configurations but not for marginally unstable or unstable ones. However, the time to reach the plateau scales as $\sim \mu^{-1} \sqrt{N}$ in contrast to the survival probabilities. The entanglement entropies closely resemble the survival probabilities in times scales, and, for unstable configurations reach the maximal values $\approx \log_2(N/3)$. 
 
We have also analyzed the evolution of the entropy using a correspondence between the survival probability and the entanglement entropy in a beam obtained using a semi-classical approximation employing Holstein-Primakoff approximation. This correspondence shows directly that flavor evolution in our system is necessarily accompanied by an increase of the entanglement entropy. The good agreement obtained between this approach and the exact numerical simulations suggests that semi-classical approaches might provide a powerful tool to explore neutrino dynamics in large systems for short time-scales. This will be especially interesting in more complicated situations with a large number of neutrino beams where the angular momentum basis scheme employed here will become computationally too expensive and the entanglement entropy might become too large for tensor network simulations. Finally, simulations using quantum devices~\cite{Hall:2021,yeter2022collective,illa2022basic} will likely become important in order to study the long time evolution of these systems.

%\newpage

%========================================================================================
\begin{acknowledgements}
%========================================================================================

ER acknowledges the NSF N3AS Physics Frontier Center, NSF Grant No. PHY-2020275, and the Heising-Simons Foundation (2017-228).  ZX was supported by the European Research Council (ERC) under the European Union's Horizon 2020 research and innovation programme (ERC Advanced Grant KILONOVA No.~885281). The work of AR was supported in part by the U.S. Department of Energy, Office of Science, Office of Nuclear Physics, Inqubator for Quantum Simulation (IQuS) under Award Number DOE (NP) Award DE-SC0020970. Calculations were carried out at the Minnesota Supercomputing Institute.
\end{acknowledgements}

\bibliographystyle{apsrev}
\bibliography{references}
%========================================================================================

\appendix

\section{Three beam geometry}
\label{app:3beams}

We assume the neutrino system is comprised of three beams and the neutrinos in each beam are parallel. This makes for a total of 3 different directions: $\vec{A},\vec{B},\vec{C}$. These vectors form a tetrahedron with volume,
\begin{equation*}
 V =\frac{A B C}{6}\sqrt{1+2 c_{A B} c_{A C} c_{B C}-c_{A B}^2-c_{A C}^2-c_{B C}^2}.
\end{equation*}
where, $c_{A B}$ is the cosine of the angle between vectors $\vec{A}$ and $\vec{B}$.
The non-negativity of the volume requires
\begin{equation*}
 1+2 c_{A B} c_{A C} c_{B C}-c_{A B}^2-c_{A C}^2-c_{B C}^2 \geq 0.
\end{equation*}
For a given volume, two of the three cosines are free parameters. 
To further simplify our analysis, we assume the three vectors are coplanar and two are antiparallel,
\begin{equation*}
 c=c_{AC}=-c_{BC},\ c_{AB}=-1
\end{equation*}
The corresponding Hamiltonian becomes
\begin{equation}
\begin{split}
H_{ABC} &= \mu\frac{4}{N}\vec{J}_A\cdot\vec{J}_B+2\mu\frac{1-c}{N}\vec{J}_A\cdot\vec{J}_C\\
&+2\mu\frac{1+c}{N}\vec{J}_B\cdot\vec{J}_C\;,
\end{split}
\end{equation}
where $N=N_A+N_B+N_C$ is the total number of spins. Note that we used the fact that $\vec{J}^2_{A_i}$ is conserved for each one of the beams.

 \section{The method in angular momentum representation}
 \label{app:angular_momentum_method}
 
 The equations of motion for the amplitudes of the many-body state defined in Eq.~\ref{eq:total_wavefunction_2beam} is
\begin{align}
    & i \partial_t a_{m_A, m_B} 
    = T_{m_A, m_B}^{m_A, m_B} a_{m_A, m_B} \nonumber\\
    + & T_{m_A,m_B}^{m_A+1, m_B} a_{m_A+1, m_B}
    + T_{m_A,m_B}^{m_A, m_B+1} a_{m_A, m_B+1} \nonumber\\
    + & T_{m_A,m_B}^{m_A+1, m_B-1} a_{m_A+1, m_B-1}
    + T_{m_A,m_B}^{m_A-1, m_B+1} a_{m_A-1, m_B+1} \nonumber\\
    + & T_{m_A,m_B}^{m_A-1, m_B} a_{m_A-1, m_B}
    + T_{m_A,m_B}^{m_A, m_B-1} a_{m_A, m_B-1},
    \label{eq:EoM_3beam}
\end{align}
where
\begin{align}
    T_{m_A, m_B}^{m_A, m_B}
    = & \frac{\mu J_{AC}}{N} [k_C (N_A-k_A) + k_A (N_C-k_C)] \nonumber\\
    & + \frac{\mu J_{BC}}{N} [k_C (N_B-k_B) + k_B (N_C-k_C)] \nonumber \\
    & + \frac{\mu J_{AB}}{N} [k_A k_B + (N_A-k_A) (N_B-k_B)] , \nonumber\\
    T_{m_A,m_B}^{m_A+1, m_B} = & T_{m_A+1, m_B}^{m_A,m_B} \nonumber\\
    = & \frac{\mu J_{AC}}{N} \, \sqrt{k_A k_C (N_A-k_A+1) (N_C-k_C+1)} , \nonumber\\
    T_{m_A,m_B}^{m_A, m_B+1} = & T_{m_A, m_B+1}^{m_A,m_B} \nonumber\\
    = & \frac{\mu J_{BC}}{N} \, \sqrt{k_B k_C (N_B-k_B+1) (N_C-k_C+1)} , \nonumber\\
    T_{m_A,m_B+1}^{m_A+1, m_B} = & T_{m_A+1, m_B}^{m_A,m_B+1} \nonumber\\
    = & \frac{\mu J_{AB}}{N} \, \sqrt{k_A k_B (N_A-k_A+1) (N_B-k_B+1)},
    \label{eq:tranelem_3beam}
\end{align}
and $k_A$, $k_B$, and $k_C$ are the flipping numbers with $m_A = N_A/2- k_A$, $m_B = N_B/2-k_B$, and $m_C = k_C-N_C/2$ respectively.
The polarization is related to the projection of flavor isospin
\begin{equation}
    \mathcal P_{A_i} = 2 \langle J^z_{A_i} \rangle/N_{A_i}
    = \sum_{m_A, m_B} \frac{2 m_A}{N_A} |a_{m_A, m_B}|^2,
    % \label{eq:P_3beam}
\end{equation}
and the pair correlations are
\begin{align}
    \langle J^x_A J^x_A \rangle = 
    \frac{1}{4} \sum_{m_A, m_B} & (N_A +2 k_A N_A -2 k_A^2) |a_{m_A, m_B}|^2,
    \nonumber\\
    \langle J^x_A J^x_B \rangle = 
    \frac{1}{2} \sum_{m_A, m_B} & \sqrt{k_A k_B (N_A-k_A+1)(N_B-k_B+1)}  \nonumber\\
    & \times Re(a_{m_A,m_B+1}^* a_{m_A+1,m_B}), \nonumber\\
    \langle J^x_A J^x_C \rangle = 
    \frac{1}{2} \sum_{m_A, m_B} & \sqrt{k_A k_C (N_A-k_A+1)(N_C-k_C+1)}  \nonumber\\
    & \times Re(a_{m_A,m_B}^* a_{m_A+1,m_B}).
    % \label{eq:Cxx_3beam}
\end{align}

 \section{Semi-classical expansion}
\label{app:hp}
In this section we introduce the truncated Holstein-Primakoff transformation, already used in~\cite{Lerose:2018,Lerose:2020} for spin systems with long-range interactions, and show how pair correlation in the neutrino beams are directly connected with the entanglement in the system.

As a first step we introduce canonical bosonic operators $p_i$ and $q_i$ for each beams as follows
\begin{equation}
\label{eq:hp_spin_basis}
\left\{\begin{matrix}
J_{A_i}^x = \sqrt{\frac{N_{A_i}}{2}}q_i + \mathcal{O}\left(\frac{1}{\sqrt{N_{A_i}}}\right)\\
J_{A_i}^y = \sqrt{\frac{N_{A_i}}{2}}p_i + \mathcal{O}\left(\frac{1}{\sqrt{N_{A_i}}}\right)\\
\pm_i J_{A_i}^z=\frac{N_{A_i}}{2}-\frac{q^2_i+p_i^2-1}{2}\\
\end{matrix}\right.\;,
\end{equation}
where the symbol $\pm_i$ denotes a $+$ sign for beams that started in the $e$ flavor (positive z polarization) and a $-$ sign for beams that started in the $x$ flavor (negative z polarization).
Note that the commutation relations of the spin operators are preserved only in the asymptotic regime $N_{A_i}\gg1$ for which $J_{A_i}^z\approx N_{A_i}/2$. This approximation is useful around the limit for which the number of excitations measured by the operator
\begin{equation}
\hat{n}_i=\frac{q^2_i+p_i^2-1}{2}
\end{equation}
remains small compared to $N_{A_i}$, a condition that for our system is fulfilled with good accuracy only for stable solutions. We will also approximate the state of each beam as a Gaussian state with covariance matrix
\begin{equation}
G_{A_i} = \begin{pmatrix}
\langle q_i^2\rangle &\frac{\langle q_ip_i+p_iq_i\rangle}{2}\\
\frac{\langle q_ip_i+p_iq_i\rangle}{2} & \langle p_i^2\rangle\\
\end{pmatrix}\;.
\end{equation}
Since we start from a product state we expect this approximation to hold for sufficiently short evolution times. Following the construction in~\cite{Lerose:2020}, we will use this approximation for the beam wave-functions to compute an approximation to the entanglement entropy of each beam. The result reads
\begin{equation}
\begin{split}
S_{A_i}(t) &= \frac{2}{\log(2)}\sqrt{\text{det}G_{A_i}(t)}\text{arccoth}\left(2\sqrt{\text{det}G_{A_i}(t)}\right)\\
&+\frac{1}{2}\log_2\left(\text{det}G_{A_i}(t)-\frac{1}{4}\right)\;,
\end{split}
\end{equation}
where we made explicit the time dependence of the covariance matrix, and thus the entropy. The covariance matrix completely characterizes the entanglement properties of a Gaussian state and we can therefore also compute other entanglement measures such as the R\'{e}nyi 2 entropy (defined in Eq.~\eqref{eq:renyi_entropy_gen} of the main text)
\begin{equation}
\begin{split}
R_{2,i}(t) &= \log_2\left(2\sqrt{\text{det}G_{A_i}(t)}\right)\\
&=1+\frac{1}{2}\log_2\left(G_{A_i}(t)\right)\;.
\end{split}
\end{equation}

In order to calculate the determinant, we first rewrite the diagonal element in terms of spin operators
\begin{equation}
\begin{split}
\langle q_i^2\rangle &= \frac{2}{N_{A_i}}\langle J_{A_i}^xJ_{A_i}^x\rangle\quad
\langle p_i^2\rangle = \frac{2}{N_{A_i}}\langle J_{A_i}^yJ_{A_i}^y\rangle\;,
\end{split}
\end{equation}
due to the $U(1)$ symmetry shared by both the initial state and the Hamiltonian these expectation values remain equal at all times. For the off-diagonal terms instead, we first introduce ladder operators
\begin{equation}
J_{A_i}^{\pm} = J_{A_i}^x\pm iJ_{A_i}^y = \sqrt{\frac{N_{A_i}}{2}}\left(q_i\pm ip_i\right)\;,
\end{equation}
from which we find
\begin{equation}
\frac{\langle q_ip_i+p_iq_i\rangle}{2} = -\frac{i}{N_{A_i}}\langle J_{A_i}^+J_{A_i}^+-J_{A_i}^-J_{A_i}^-\rangle\;,
\end{equation}
this are also zero for our system due to the conservation of the total spin. We can now proceed in two ways: the first one is to use the definition of $J_{A_i}^z$ in Eq.~\eqref{eq:hp_spin_basis} to write
\begin{equation}
\begin{split}
\langle q_i^2\rangle = \langle p_i^2\rangle &= \frac{1}{2}+\langle \hat{n}_i\rangle\\
&= \frac{1}{2}+\frac{N_{A_i}}{2}\mp_i\langle J_{A_i}^z\rangle\\
&= \frac{1}{2}+N_{A_i}\left(1-P_{A_i}(t)\right)\;,
\end{split}
\end{equation}
where in the last line we avoided the beam-dependent $\mp_i$ sign by using the definition of flavor survival probability for beam $i$ from Eq.~\eqref{eq:surv_prob} of the main text (note that here we haven't indicated the initial condition).
We can express the covariance matrix as
\begin{equation}
G^{(a)}_i= \left(\frac{1}{2}+N_{A_i}\left(1-P_{A_i}(t)\right)\right)\begin{pmatrix}
1&0\\
0&1\\
\end{pmatrix}\;.
\end{equation}
The second one is to use the conservation of the angular momentum $J_{A_i}^2$ to write the covariance matrix as
\begin{equation}
\begin{split}
G_i^{(b)}&=\frac{2}{N_{A_i}}\begin{pmatrix}
\langle J_{A_i}^xJ_{A_i}^x\rangle&0\\
0&\langle J_{A_i}^yJ_{A_i}^y\rangle\\
\end{pmatrix}\\
&=\frac{J_{A_i}^2-\langle J_{A_i}^zJ_{A_i}^z\rangle}{N_{A_i}}\begin{pmatrix}
1&0\\
0&1\\
\end{pmatrix}\\
&=\left(\frac{1}{2}+\frac{N_{A_i}}{4}-\frac{\langle J_{A_i}^zJ_{A_i}^z\rangle}{N_{A_i}}\right)\begin{pmatrix}
1&0\\
0&1\\
\end{pmatrix}\;,\end{split}
\end{equation}
where in the last step we have used the initial value $J_{A_i}^2=N_{A_i}(N_{A_i}+2)/4$ valid for every beam.

We finally find the following compact expression for the Von Neumann entropy in both approximations as
\begin{equation}
\begin{split}
S^{(a/b)}_{A_i}(t) &= \frac{1+2\Gamma^{(a/b)}_{A_i}(t)}{\log(2)}\text{arccoth}\left(1+2\Gamma^{(a/b)}_{A_i}(t)\right)\\
&+\frac{1}{2}\left[\log_2(\Gamma^{(a/b)}_{A_i}(t))+\log_2(1+\Gamma^{(a/b)}_{A_i}(t))\right]\;,
\end{split}
\end{equation}
and, correspondingly, the R\'{e}nyi $2$ entropy becomes
\begin{equation}
R_{2,A_i}(t) = \log_2\left(1+2\Gamma^{(a/b)}_{A_i}(t)\right)\;.
\end{equation}
In these expression we have introduced the quantity
\begin{equation}
\Gamma^{(a)}_{A_i}(t) = N_{A_i}\left(1-P_{A_i}(t)\right)
\end{equation}
for approximation $(a)$ and
\begin{equation}
\Gamma^{(b)}_{A_i}(t) = \frac{N_{A_i}}{4}-\frac{\langle J_{A_i}^zJ_{A_i}^z\rangle}{N_{A_i}}\;,
\end{equation}
for approximation $(b)$. At the beginning of time evolution $\Gamma^{(a/b)}_i(t)=0$ and so is the entropy. The largest value this can reach in approximation $(a)$ is when the survival probability goes to zero while in approximation $(b)$ when all the angular momentum is in the $(X,Y)$ plane and $\langle J_{A_i}^zJ_{A_i}^z\rangle=0$. In these limits the entropy is approximately
\begin{equation}
\begin{split}
S^{(a)}_{max} \approx 1+\log_2\left(N_{A_i}\right)\\
S^{(b)}_{max} \approx 1+\log_2\left(\frac{N_{A_i}}{4}\right)\;.
\end{split}
\end{equation}
Since the $(b)$ approximation depends directly on the approximate definition of the spin operators in the $X$ and $Y$ direction from Eq.~\eqref{eq:hp_spin_basis}, we expect it to break down when $\langle J^z_{A_i}\rangle$ deviates significantly from its initial value. On the other hand  approximation $(a)$ only relies on this mapping to establish $\langle q_i^2\rangle=\langle p_i^2\rangle$ and that the off diagonal elements of $G_i$ are zero but is otherwise exact (within the Gaussian approximation). We then expect approximation $(a)$ to perform better in practice in the limit of large system size.

\end{document}